\documentclass[useAMS,usenatbib]{mn2e}

\usepackage{graphicx,amssymb,amstext} 
\usepackage{hyperref}
\usepackage[T1]{fontenc}
\usepackage{aecompl}
 
\def\aj{AJ}			
\def\araa{ARA\&A}		
\def\apj{ApJ}		
\def\apjl{ApJ}		
\def\apjs{ApJS}				
\def\apss{Ap\&SS}		
\def\aap{A\&A}

\def\mnras{MNRAS}

\begin{document}

\title[The co-evolution of black hole growth and star formation]{The co-evolution of black hole growth and star formation from a cross-correlation analysis between quasars and the cosmic infrared background}
\author[L.~Wang et al.]
{\parbox{\textwidth}{\raggedright Lingyu Wang$^{1,}$\thanks{E-mail: \texttt{lingyu.wang@durham.ac.uk}}, Marco Viero$^{2}$, Nicholas P. Ross$^{3}$,  Viktoria Asboth$^{4}$, Matthieu B{\'e}thermin$^{5}$, Jamie Bock$^{2, 6}$, Dave Clements$^{7}$, Alex Conley$^{8}$, Asantha Cooray$^{9}$, Duncan Farrah$^{10}$, Amir Hajian$^{11}$,  Jiaxin Han$^{1}$,  Guilaine Lagache$^{12}$, Gaelen Marsden$^{4}$,  Adam Myers$^{13}$, Peder Norberg$^{1}$,  Seb Oliver$^{14}$, Mat Page$^{15}$, Myrto Symeonidis$^{14}$, Bernhard Schulz$^{6, 16}$,  Wenting Wang$^{1}$, Mike Zemcov$^{2, 6}$}\vspace{0.4cm}\\
\parbox{\textwidth}{\raggedright $^{1}$Institute for Computational Cosmology, Department of Physics, Durham University, Durham, DH1 3LE, UK\\
$^{2}$California Institute of Technology, 1200 E. California Blvd., Pasadena, CA 91125\\
$^{3}$Department of Physics, Drexel University, 3141 Chestnut Street, Philadelphia, PA 19104, USA\\
$^{4}$Department of Physics \& Astronomy, University of British Columbia, 6224 Agricultural Road, Vancouver, BC V6T 1Z1, Canada\\
$^{5}$European Southern Observatory, Karl-Schwarzschild-Str. 2, 85748 Garching, Germany\\
$^{6}$Jet Propulsion Laboratory, 4800 Oak Grove Drive, Pasadena, CA 91109\\
$^{7}$Astrophysics Group, Blackett Laboratory,  Imperial College of Science Technology and Medicine, London SW7 2BZ, UK\\
$^{8}$Center for Astrophysics and Space Astronomy 389-UCB, University of Colorado, Boulder, CO 80309\\
$^{9}$Center for Cosmology, Department of Physics and Astronomy, University of California, Irvine, CA 92697\\
$^{10}$Department of Physics, Virginia Tech, Blacksburg, VA 24061, USA\\
$^{11}$Canadian Institute for Theoretical Astrophysics, University of Toronto, Toronto, ON M5S 3H8, Canada\\
$^{12}$Institut d'Astrophysique Spatiale (IAS), B\^atiment 121, F- 91405 Orsay (France); Universit\'e Paris-Sud 11 and CNRS (UMR 8617)\\
$^{13}$Department of Physics and Astronomy, University of Wyoming, Laramie, WY 82071, USA\\
$^{14}$Astronomy Centre, Dept. of Physics \& Astronomy, University of Sussex, Brighton BN1 9QH, UK\\
$^{15}$Mullard Space Science Laboratory, University College London, Holmbury St. Mary, Dorking, Surrey RH5 6NT, UK\\
$^{16}$Infrared Processing and Analysis Center, MS 100-22, California Institute of Technology, JPL, Pasadena, CA 91125}}

\date{Accepted . Received ; in original form }

\maketitle

\begin{abstract}
We present the first cross-correlation measurement between Sloan Digital Sky Survey (SDSS) Type 1 quasars and the cosmic infrared background (CIB) measured by {\em Herschel}-SPIRE. The distribution of the quasars at $0.15<z<3.5$ covers the redshift range where we expect most of the CIB to originate. We detect the sub-millimetre (sub-mm) emission of the quasars, which dominates on small scales, as well as correlated emission from dusty star-forming galaxies (DSFGs) dominant on larger scales.  A simple halo model is used to interpret the measured cross-correlation signal. The mean sub-mm flux densities of the DR7 quasars (median redshift  $\left<z\right>=1.4$) is $11.1^{+1.6}_{-1.4}$, $7.1^{+1.6}_{-1.3}$ and $3.6^{+1.4}_{-1.0}$ mJy at 250, 350 and 500\ $\micron$, respectively, while the mean sub-mm flux densities of the DR9 quasars ($\left<z\right>=2.5$) is $5.7^{+0.7}_{-0.6}$, $5.0^{+0.8}_{-0.7}$ and $1.8^{+0.5}_{-0.4}$ mJy. We find that the correlated sub-mm emission  includes both the emission from satellite DSFGs in the same halo as the central quasar (the one-halo term) and the emission from DSFGs in separate halos that are correlated with the quasar-hosting halo (the two-halo term). The amplitude of the one-halo term is about 10 times smaller than the sub-mm emission of the quasars, implying the the satellite DSFGs have a lower star-formation rate than the central quasars. We infer that the satellite fraction (i.e. the fraction of quasars hosted by satellite galaxies in the halo) for the DR7 quasars  ($\left<z\right>=1.4$) is $0.008^{+0.008}_{-0.005}$ and the host halo mass scale for the central and satellite quasars is $10^{12.36\pm0.87}$ M$_{\odot}$ and $10^{13.60\pm0.38}$ M$_{\odot}$, respectively. At a median redshift of 2.5,  the satellite fraction of the DR9 quasars is $0.065^{+0.021}_{-0.031}$ and the host halo mass scale for the central and satellite quasars is $10^{12.29\pm0.62}$ M$_{\odot}$ and $10^{12.82\pm0.39}$ M$_{\odot}$, respectively.  Thus, the typical halo environment of the SDSS Type 1 quasars is found to be similar to that of DSFGs over the redshift range probed, which supports the generally accepted view that dusty starburst and quasar activity are evolutionarily linked phenomena.

\end{abstract}


\section{INTRODUCTION}

Quasars, powered by accretion of material onto the central supermassive black holes (SMBHs) of active galaxies, are among the most luminous objects in the Universe. In the current paradigm of galaxy formation and evolution, quasars represent a key phase in the evolutionary history of massive galaxies. Essentially all spheroidal systems are believed to harbour massive black holes. The masses of the central SMBHs are shown to correlate with many properties of their host galaxies, e.g., the stellar velocity dispersion of the bulge, the bulge mass and the bulge luminosity (e.g., Kormendy \& Richstone 1995; Magorrian et al. 1998; Ferrarese \& Merritt 2000; Gebhardt et al. 2000; Tremaine et al. 2002; Kormendy et al. 2011). Present-day SMBHs are thought to have gained most of their mass via accretion during an active nuclear phase, e.g. the quasar phase. The implication is that black hole growth and star-formation activity are inextricably linked, at least for galaxies which experience `wet' major mergers and a quasar-like period of rapid black hole growth (see Kormendy \& Ho 2013 for a review). A probable scenario is that the galaxy initially forms in a gas-rich rotationally supported system. As the host dark matter halo grows, some mechanisms, such as major merger or disk/bar instability trigger a period of rapid, dust obscured star-formation activity (starburst) which produces a stellar bulge. At the same time, the infalling gas fuelling the starburst is also feeding the central black hole. After some time the quasar becomes optically visible (unobscured), and soon after that the on-going star formation is quenched on a short time-scale, perhaps via radiative or mechanical feedback from the central engine (see Fabian 2012 for a review). The cosmic star-formation history is shown to be similar to the evolving luminosity density of quasars (which is related to the BH accretion history) at most redshifts (e.g., Boyle \& Terlevich 1998; Franceschini et al. 1999; Merloni et al. 2004; Shankar et al. 2007; Silverman et al. 2008; Wall, Pope \& Scott 2008), which adds further support to this picture of the galaxy - black hole connection.

In this paper, we probe the co-evolution of black hole growth and star-formation activity using a cross-correlation analysis between optically selected Type 1 quasars over the redshift range $0.2<z<3.5$ and the cosmic infrared background (CIB) in the far-infrared (FIR)/sub-millimetre (sub-mm). The former represents some of the most luminous and massive black holes found to date. The latter is dominated by emission from dusty star-forming galaxies (DSFGs) over a similar redshift range to the quasars (Valiante et al. 2009; Bethermin et al. 2011; Bethermin et al. 2012; Viero et al. 2013a). The large-scale cross-correlation signal of quasars and star-forming galaxies is determined by the effective linear bias (or equivalently, the effective host halo mass) of both sets of tracers of the underlying large-scale structure. The small-scale cross-correlation signal constrains the halo occupation distributions (HODs) of star-forming galaxies in quasar-hosting dark matter halos and the HODs of quasars in halos hosting star-forming galaxies. In contrast, the auto-correlation function constrains the HODs of each population independently, not the HODs of one population in the presence of the other. Therefore, the cross-correlation function is arguably more suited to studying the co-evolution of the quasars and the star-forming galaxies\footnote{Ideally, one can combine the cross- and auto-correlation functions to constrain the HODs of different galaxy populations. In practice, however, the auto-correlation function is more difficult to measure accurately because it is very sensitive to how well we can understand and reproduce the selection effects of the sample. In contrast, uncorrelated selection effects only affect the uncertainty but do not bias measurement of the cross-correlation signal.}. 


Quasars are very luminous active galactic nuclei (AGN) and hence can be seen out to great distances. They are great tracers of the large-scale structure. To measure their clustering properties reliably, large-volume surveys are required, due to the low spatial density of quasars. The Sloan Digital Sky Survey (SDSS; York et al. 2000) has produced some of the largest quasar samples to date and that is what we use in this paper. 

FIR and sub-mm observations are critical for studying star-forming galaxies, as dust absorbs the optical/ultraviolet light and reradiates in the FIR/sub-mm. Dust emission makes up the CIB, which contains about half of the extragalactic background radiation (excluding the cosmic microwave background), i.e. the total power originated from stars and AGN throughout the cosmic history (Puget et al. 1996; Fixsen et al. 1998; Lagache et al. 1999). The \emph{Herschel\footnote{\emph{Herschel} is an ESA space observatory with science instruments provided by European-led Principal Investigator consortia and with important participation from NASA.} Space Observatory} (Pilbratt et al. 2010) has carried out imaging surveys with unprecedented size and depth at FIR/sub-mm wavelengths. Thanks to its sensitivity and mapping speed, it allows us for the first time to probe the bolometric power of DSFGs at high redshifts in large volumes. 

This paper is organised as follows. In Section 2, we give an overview of the \emph{Herschel}-SPIRE maps and SDSS DR7 and DR9 QSO catalogues used in our analysis. In Section 3, we first describe the estimator used to calculate the cross-correlation signal between the QSOs and the CIB. We then investigate the potential contamination of Galactic dust emission by explicitly calculating the cross-correlation signal between our quasar samples and the IRAS 100\ $\micron$ data in our fields. Finally, we present the cross-correlation results between various quasar samples and the SPIRE maps at 250, 350 and 500\ $\micron$.  In Section 4, a simple halo model is employed to interpret the measured QSO -- CIB cross-correlation signal and constrain the HODs of quasars in halos that also host star-forming galaxies. Further discussion and conclusions of our results are given in Section 5. We assume an $H_0=70$ km s$^{-1}$ Mpc$^{-1}$, $\Omega_{\rm M}=0.3$, and $\Omega_{\Lambda}=0.7$ cosmology and use the AB magnitude scale throughout the paper.

\section{Data}

\begin{figure*}
\includegraphics[height=5.5in,width=7.in]{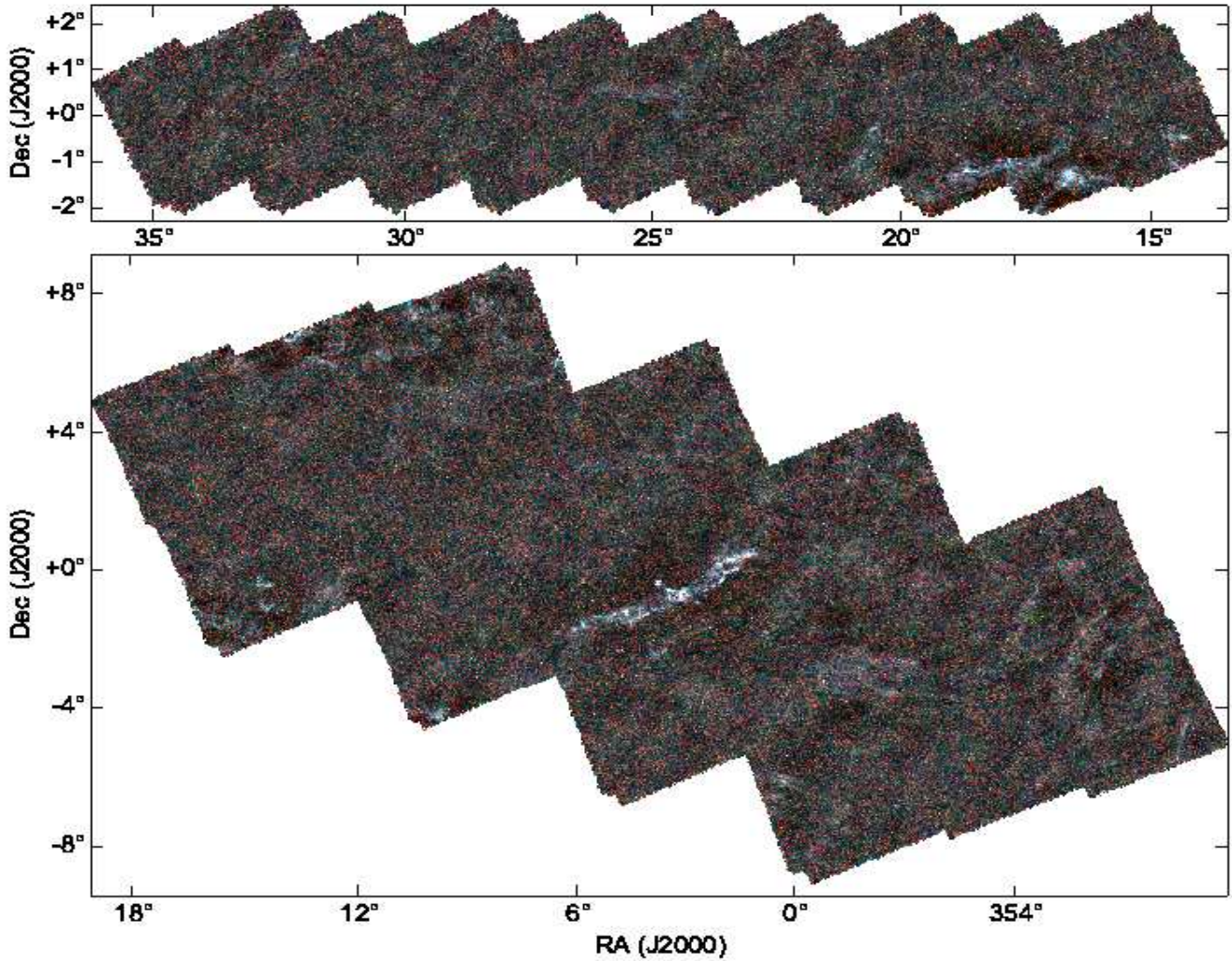}
\caption{Three-colour SPIRE images of the HeRS (top) and HeLMS (bottom) fields with 250, 350, and 500\ $\micron$ as blue, green, and red, respectively. Foreground clouds of Galactic cirrus, which correlate with HI emission, can be clearly seen by eye. The areal coverage of HeRS and HeLMS is 79 deg$^2$ and 270 deg$^2$, respectively. Jointly, they cover the full $\sim$ 150 deg$^2$ subset of the SDSS Stripe 82 region that has the lowest level of Galactic cirrus foregrounds. }
\label{fig:hershelms_sky}
\end{figure*}

\subsection{\emph{Herschel}-SPIRE maps}

We use  CIB maps at  250, 350, and 500 $\micron$ (1200, 857, and 600\,GHz, respectively) from the two \emph{Herschel} surveys in the SDSS Stripe 82 region, i.e. the \emph{Herschel} Stripe 82 Survey (HerS\footnote{HeRS maps and point source catalogues can be accessed from \url{http://www.astro.caltech.edu/hers/.}}; Viero et al. 2013b) and the HerMES\footnote{The \emph{Herschel} Multi-tiered Extragalactic Survey (HerMES; \url{http://hermes.sussex.ac.uk}) is a guaranteed time key project of the \emph{Herschel Space Observatory}. HerMES maps and point source catalogues are publicly available at \url{http://hedam.lam.fr/HerMES}.} Large-Mode Survey (HeLMS; Oliver et al. 2012). Maps were observed with the Spectral and Photometric Imaging Receiver (SPIRE) instrument (Griffin et al. 2010) aboard \emph{Herschel}. HeRS covers 79 deg$^2$ along the SDSS Stripe 82 to a 5-$\sigma$ depth of 65.0,  64.5 and  74.0 mJy beam$^{-1}$ at 250, 350 and 500\ $\micron$, respectively. HeLMS covers 270 deg$^2$ to a a 5-$\sigma$ depth of 64.0, 53.0 and 76.5 mJy beam$^{-1}$ at 250, 350 and 500\ $\micron$. The joint HeRS and HeLMS areal coverage between -10$^{\circ}$ and 37$^{\circ}$ (RA) covers the full $\sim 150\, \rm deg^2$ subset of Stripe 82 that has the lowest level of Galactic dust emission (or cirrus) foregrounds (i.e., with $N_{\rm H}\lesssim 2\times 10^{21}\, \rm \, cm^{-2}$).

The SPIRE data, obtained from the \emph{Herschel} Science Archive, were reduced using the standard ESA software and the custom-made software package, {\sc SMAP} (Levenson et al. 2010, Viero et al. 2013b). Maps were made using an updated version of SMAP/SHIM, which is an iterative map-maker designed to optimally separate large-scale noise from signal. For more details, please refer to Viero et al. (2013b). The maps have a tangent plane (TAN) projection with pixel sizes of 6$\arcsec$, 8.33$\arcsec$ and 12$\arcsec$ at  250, 350, and 500 $\micron$, respectively. In comparison, the full width at half maximum (FWHM) of the SPIRE beams are 18.1$\arcsec$, 25.2$\arcsec$ and 36.6$\arcsec$ at  250, 350, and 500 $\micron$, respectively (Swinyard et al. 2010). The SPIRE maps are converted from their native unit of $\rm Jy\, beam^{-1}$ to $\rm Jy\, sr^{-1}$ by dividing them by the effective beam areas which are 0.9942, 1.765, and 3.730 $\times 10^{-8}$\,steradians at  250, 350, and 500 $\micron$, respectively (SPIRE Observers' Manual\footnote{\url{http://herschel.esac.esa.int/Docs/SPIRE/html/spire_handbook.html}}).  We note that no colour corrections from a flat-spectrum point-source calibration are applied, as they were shown in Viero et al. (2013a) to be negligible. Fig.~\ref{fig:hershelms_sky} shows the three-colour SPIRE images of the HeRS and HeLMS regions. We can clearly see white foreground clouds of Galactic cirrus emission, which have been shown to correlate with HI emission from GASS 21-cm data in Viero et al. (2013b).

\begin{figure}
\includegraphics[height=8.9in,width=3.35in]{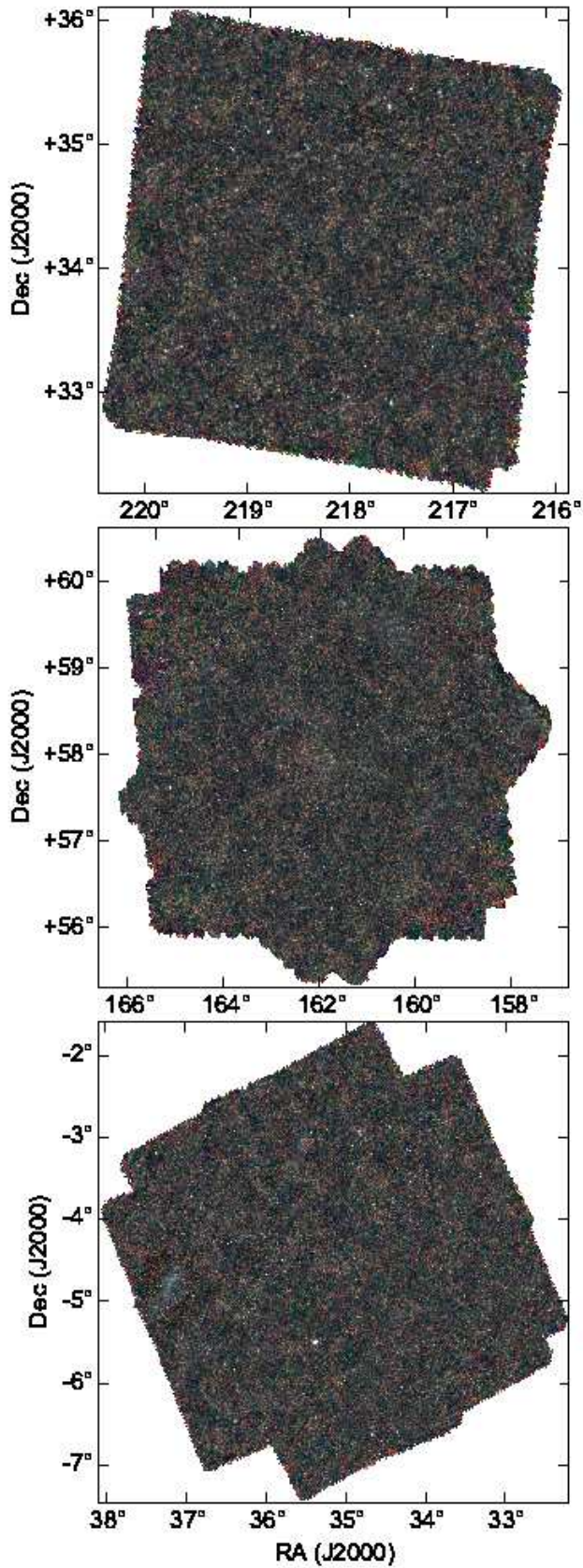}
\caption{Three-colour SPIRE images of the 11.3 deg$^2$ Bo\"{o}tes filed (top), the 15.2 deg$^2$ Lockman-SWIRE field (middle) and the 21.6 deg$^2$ {\it XMM}-LSS field (bottom).}
\label{fig:lss14_sky}
\end{figure}

In this paper, we also include SPIRE maps of a subset of the HerMES level 5 and level 6 fields\footnote{The HerMES fields are organised, according to area and depth, into levels 1 through to 7, with level 1 fields being the smallest and deepest, and level 7 being the widest and shallowest. HeLMS is the only level 7 field.}, which have significant overlaps with our SDSS quasar samples. In particular, the HerMES Bo\"{o}tes field (11.3 deg$^2$) and the Lockman-SWIRE field (15.2 deg$^2$) overlap with the SDSS DR7 QSO sample. The Bo\"{o}tes field and the {\it XMM}-LSS field (21.6 deg$^2$) overlap with the SDSS DR9 QSO sample. Fig.~\ref{fig:lss14_sky} shows the three-colour SPIRE images of the Bo\"{o}tes, Lockman-SWIRE and {\it XMM}-LSS field.  Out of these three fields, the {\it XMM}-LSS field is the most contaminated by cirrus emission, but all three fields have significantly less cirrus contamination than the HeRS or HeLMS field. In addition, the level 5 and 6 fields are significantly deeper than HeRS or HeLMS. The 5-$\sigma$ noise level in the {\it XMM}-LSS and Bo\"{o}tes field is 25.8, 21.2 and 30.8 mJy at 250, 350 and 500\ $\micron$, respectively. The Lockman-SWIRE field is even deeper and the 5-$\sigma$ noise level is 13.6, 11.2 and 16.2 mJy at 250, 350 and 500\ $\micron$, respectively.

The CIB at sub-mm wavelengths is dominated by emission from DSFGs over a wide range of redshift, while other potential sources of signal, e.g., the cosmic microwave background (CMB), Sunyaev-Zel'dovich (SZ) effect, and radio galaxies, are subdominant. One exception is the emission from Galactic dust (cirrus) which cannot be ignored at FIR/sub-mm wavelengths. Therefore, the SPIRE map signal can be modelled, to first order, as the sum of the following three terms: the emission from DSFGs; Galactic cirrus emission; and instrument noise. Thus we can write
\begin{equation}
I_{\rm sky} = I_{\rm DSFG} + I_{\rm cirrus} + I_{\rm noise}.
\end{equation}
We expect that only the emission from DSFGs have intrinsic correlation with the quasars. This is because some quasars could be strong IR/sub-mm emitters from dust heated by the accreting black hole, as well as star formation in the host galaxy (e.g. Lutz et al. 2008; Shi et al. 2009; Clements et al. 2009; Serjeant et al. 2010; Bonfield et al. 2011; Dai et al. 2012). But additionally  DSFGs are expected to occur as satellite galaxies in the dark matter halos where quasars are found\footnote{The satellite fraction of quasars (i.e. quasars hosted by satellite galaxies in the halo) is estimated to be at most a few percent (Kayo \& Oguri 2012; Richardson et al. 2012; Shen et al. 2013). Therefore, most quasars are expected to be central galaxies.}. A last effect is that dark matter halos correlated with the quasar-hosting halo are also expected to host DSFGs. The SPIRE instrumental noise should not correlate with quasars and therefore only contributes to the noise in the measured cross-correlation signal between quasars and the CIB. Similarly, Galactic dust emission should not have any intrinsic correlation with the quasars. However, due to the selection of the quasars, there might be (anti-)correlation between the observed quasar number density field and the intensity of the cirrus emission. There are two effects of Galactic dust on the selection of QSOs. The first effect is reddening which means QSOs with extreme reddening will not be selected as targets for spectroscopy. The second effect is extinction, which can cause the QSOs to drop out of the target list even before they reach the colour selection stage. QSO populations with steeper number counts will be more sensitive to extinction.

To explicitly check the potential correlation between the quasar samples and Galactic cirrus emission, we make use of the 100\ $\micron$ maps  from the Improved Reprocessing of the \textit{IRAS} (Neugebauer et al. 1984) Survey (IRIS; Miville-Deschenes \& Lagache 2005), a data set which corrects the original plates for calibration, zero level and striping problems. The IRIS 100\ $\micron$ maps in the HerMES level 5 and level 6 fields are discussed in Viero et al. (2013) and the IRIS 100\ $\micron$ maps in HeRS and HeLMS are discussed in Hajian et al. (in prep.). The 100\ $\micron$ maps in all of our fields are shown in Appendix \ref{appendix1}. Here we have assumed that the dominant signal in the 100\ $\micron$ maps comes from Galactic cirrus emission, not the DSFGs comprising the CIB. In other words, these 100\ $\micron$ maps are used as a Galactic cirrus tracer in our fields\footnote{Properties of Galactic cirrus (such as column density, dust emissivity and dust  temperature) can vary a lot from field to field. In some regions, the IRIS 100\ $\micron$ data may not be used as a reliable Galactic cirrus tracer as a result of significant contamination from the CIB (P{\'e}nin et al. 2012).}. Visual inspection of the maps in Appendix \ref{appendix1} confirms that extended diffuse Galactic cirrus emission are the dominant structures (long filamentary chains) in the 100\ $\micron$ maps, especially in the HeRS and HeLMS regions. We have also computed the mean value of the 100\ $\micron$ maps in the HeLMS, HeRS, {\it XMM}-LSS, Bo\"{o}tes and Lockman-SWIRE field to be  3.03, 2.84, 2.30, 1.42 and 1.09 MJy sr$^{-1}$, respectively. If we adopt the DIRBE measurement  of the cosmic IR background at 100\ $\micron$ which is $14.4\pm6.3$ nW m$^{-2}$ sr$^{-1}$ or equivalently $0.48\pm0.21$ MJy  sr$^{-1}$ (Dole et al. 2006), then we can estimate that the fractional contribution of the CIB to the 100\ $\micron$ maps in the HeLMS, HeRS, {\it XMM}-LSS, Bo\"{o}tes and Lockman-SWIRE field  is  16\%, 17\%, 21\%, 34\% and 44\%, respectively.

\subsection{SDSS DR7 and SDSS-III DR9 QSOs}
For our cross-correlation analysis, we use quasars from both the SDSS DR7 (Abazajian et al. 2009) and the SDSS-III DR9 (Ahn et al. 2012) spectroscopic surveys.

\subsubsection{SDSS DR7 Quasars}

The spectroscopic SDSS DR7 quasar catalogue (DR7Q; Schneider et al. 2010; Shen et al. 2010) contains 105,783 spectroscopically confirmed quasars with $i$-band absolute magnitude $M_i<-22.0$ and $0.065<z<5.46$ over an area of 9,380 deg$^2$. It is the culmination of the SDSS-I and SDSS-II quasar surveys. The quasar target selection is described in detail by Richards et al. (2002). Quasar candidates are selected based on their fluxes and colors in SDSS bands. At low-redshift, $z\lesssim3$, quasar targets are selected based on their location in $ugri$-color space and at high-redshift, $z\gtrsim3$, targets are selected from $griz$-color space. Quasar candidates passing the $ugri$-color selection are selected to a flux limit of $i=19.1$, but since high-redshift quasars are rare, objects lying in regions of color-space corresponding to quasars at $z\gtrsim3$ are targeted to $i=20.2$ (PSF magnitudes corrected for Galactic extinction based on the dust maps of Schlegel, Finkbeiner \& Davis 1998).

\subsubsection{SDSS-III BOSS DR9 Quasars}

The Data Release 9 Quasar catalogue (DR9Q; P{\^a}ris et al. 2012) is the first quasar catalogue from the Baryon Oscillation Spectroscopic Survey (BOSS; Dawson et al. 2012) of the Sloan Digital Sky Survey III (Eisenstein et al. 2011). This catalogue contains 87,822 quasars over 3,275 deg$^2$. The primary science goal for the high-$z$ BOSS quasar survey was to measure the baryon acoustic oscillations in the Lyman-$\alpha$ forest (Busca et al. 2013; Slosar et al. 2013; Delabuc et al. 2014). As such a priority was set on obtaining a high-surface density of $z>2.1$ quasars in order to maximise the number of Lyman-$\alpha$ forest lines of sight observable in the blue. To achieve this high-surface density, quasar candidates were selected in a heterogenous manner, down to a magnitude limit of  $g\leq22.0$ or $r\leq21.85$ (Ross et al. 2012). Quasars were selected based on their optical fluxes and colours, as well as their near-infrared, ultraviolet fluxes and radio properties; Ross et al. (2012; and references therein) has full details.

However, a uniform selection was performed on a subsample of the DR9 quasar targets. This uniform sample was selected by the ``Extreme Deconvolution" (XD; Bovy, Hogg \& Roweis 2009) algorithm which is applied to model the distribution in flux and flux uncertainty of a training set of quasars and stars (XDQSO; Bovy et al. 2011). The DR9Q has a flag which indicates if a quasar was selected in this uniform manner. In this paper we select all quasars in the DR9Q catalog with ``uniform $>0$" which represent a statistical sample for clustering studies (P{\^a}ris et al. 2012; White et al. 2012). Additionally, we limit the SDSS DR9 quasar sample to be in the redshift range $2.2<z<3.5$. The median redshift of the final selected SDSS DR9 quasar sample is $z=2.5$.

\begin{figure*}
\includegraphics[height=2.6in,width=3.45in]{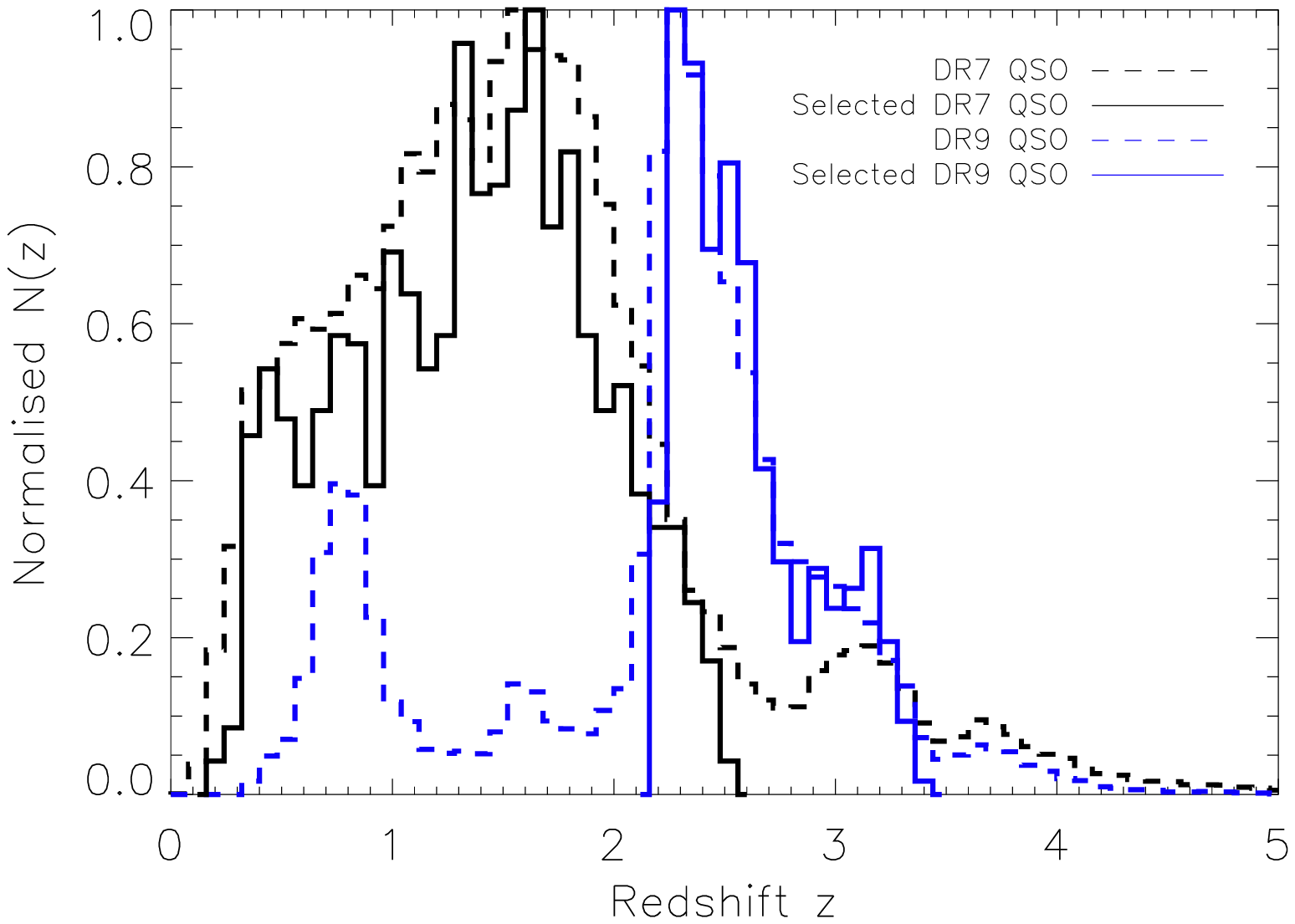}
\includegraphics[height=2.6in,width=3.45in]{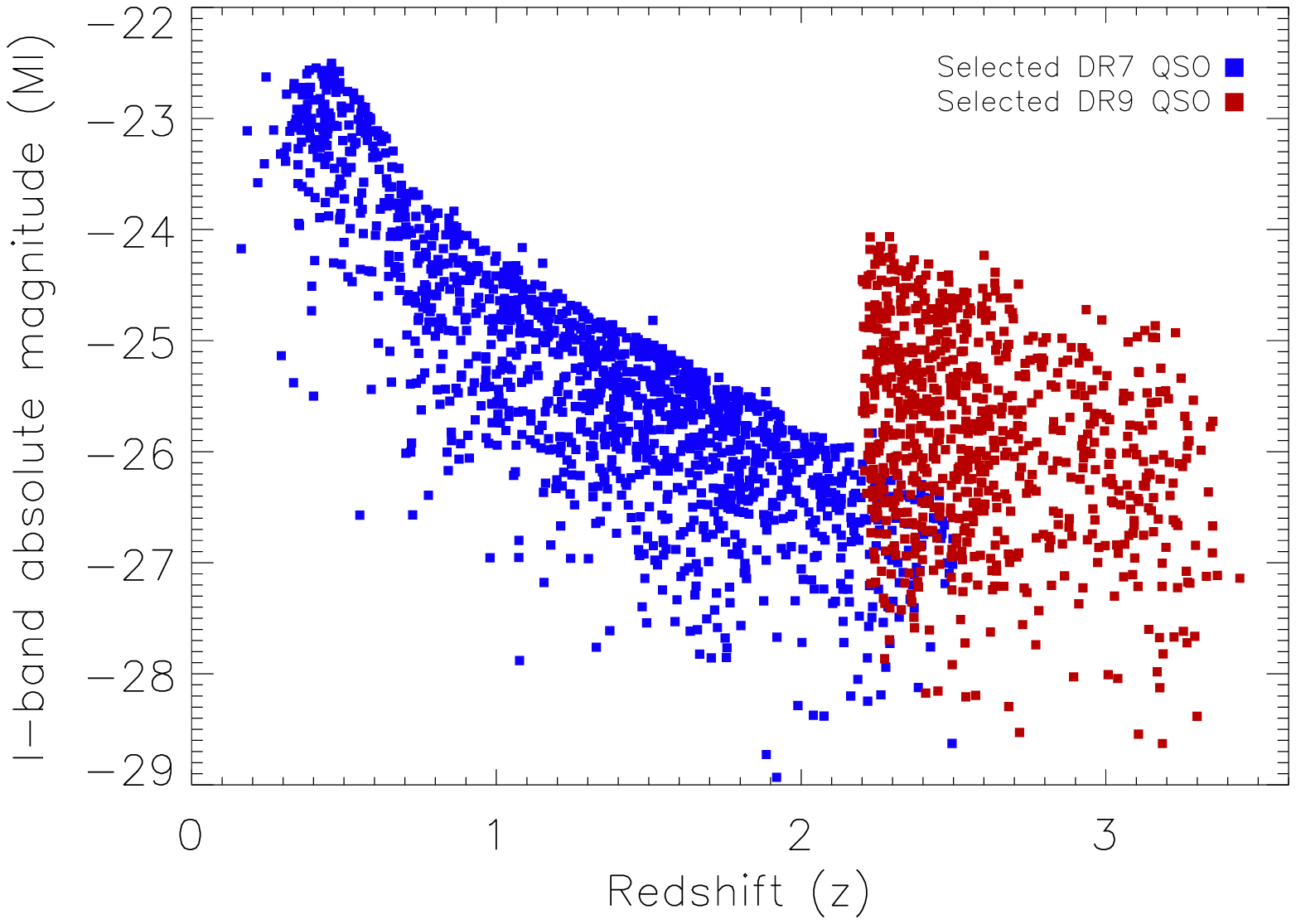}
\caption{Left: Redshift distribution of the SDSS DR7 and DR9 QSO sample. The dashed lines correspond to the normalised redshift distribution of the entire DR7 and DR9 samples, while the solid lines correspond to the selected samples used in the cross-correlation analysis. The selected DR7 QSO sample covers the redshift range $z=[0.15, 2.5]$, with a median redshift of 1.4, and the selected DR9 QSO sample covers the redshift range $z=[2.2, 3.5]$, with a median redshift of 2.5. Right: $i$-band absolute magnitude versus redshift for DR7 and DR9 quasars. The DR9 QSO sample, apart from being at higher redshift, also includes fainter objects.}
\label{fig:dndz_mi}
\end{figure*}

\subsubsection{Stripe 82}

Stripe 82 is an equatorial region that the Sloan Digital Sky Survey (SDSS) has repeatedly imaged up to 80 times (Abazajian et al. 2009) giving coadded optical data $\sim$2 magnitudes deeper than the single epoch SDSS observations. The Stripe 82 field covers approximately a 270 deg.$^2$ area ($-50^{\circ} < $RA$ < 59^{\circ}$, $-1.25^{\circ} <$ Dec $< 1.25^{\circ}$). 

For the SDSS DR7Q, since the Stripe 82 imaging is deeper we are able to select quasars to $i<19.9$ for the cross-correlation study. We further restrict our analysis to SDSS DR7 quasars in the redshift range $0.15<z<2.5$, with a median redshift of 1.4. The redshift cut is used to avoid the redshift deserts of quasars at $z\sim2.7$ and $\sim3.5$ due to contamination from F and GK stars (Fan 1999; Richards et al. 2002, 2006).

One complication for SDSS-III BOSS DR9 is that a set of quasar selection methods, and not the uniform XDQSO method, was used in the Stripe 82 region. Consequently, the White et al. (2012) completeness corrections are not suitable in Stripe 82. However, during BOSS observations, Stripe 82 was observed at a high target density (Palanque-Delabrouille et al. 2011) and the actual completeness of $2.2<z<3.5$ quasars is close to $100\%$, down to the $g\leq22.0$ or $r\leq21.85$ limiting magnitude (Palanque-Delabrouille et al. 2013; Ross et al. 2013).

\subsubsection{Quasar Redshift and Luminosity distributions}

\begin{table}
\caption{Number of selected SDSS DR7 and DR9 QSOs found in HeRS, HeLMS, Bo\"{o}tes, Lockman-SWIRE and {\it XMM}-LSS regions. The Lockman-SWIRE field does not overlap with the DR9 quasar sample and the {\it XMM}-LSS field does not overlap with the DR7 quasar sample. Note that the number of DR9 QSOs in the HeRS and HeLMS Stripe 82 regions goes through a final change in Section 3.2. The numbers inside the brackets represent the final number of DR9 quasars in these two regions. The median redshift for the DR7 and DR9 QSOs is 1.4 and 2.5, respectively.}
    \begin{tabular}{lll}
    \hline
    Region & DR7 & DR9 \\
    \hline
    HeRS (79 deg$^2$) & 1,411 & 630 (762)\\
    HeLMS (270 deg$^2$)& 1,675 & 769 (947)\\
     Bo\"{o}tes (11.3 deg$^2$)& 100 & 183 \\
     Lockman-SWIRE (15.2 deg$^2$)& 153 & - \\
     {\it XMM}-LSS (21.6 deg$^2$)& - & 117 \\
    \hline
    \end{tabular}
\end{table}

In the left panel in  Fig.~\ref{fig:dndz_mi}, we plot the normalised redshift distribution of the SDSS DR7 and DR9 QSO samples. The dashed lines correspond to the normalised redshift distribution of the entire DR7 and DR9 QSO samples, while the solid lines correspond to the selected QSO samples used in the cross-correlation analysis. Compared to the expected redshift distribution of the integrated sub-mm emission at 250, 350 and 500 $\micron$ from various models (Valiante et al. 2009; Bethermin et al. 2011; Bethermin et al. 2012; Viero et al. 2013), we can see that the SDSS DR7 and DR9 quasars completely cover the redshift range where the bulk of the sub-mm emission is expected to arise. In the right panel in  Fig.~\ref{fig:dndz_mi}, we plot the $i$-band absolute magnitude versus redshift for the DR7 and DR9 QSO sample. The DR9 QSO sample, apart from being at higher redshift, also extends to fainter objects. 

\begin{figure*}
\includegraphics[height=3.3in,width=3.45in]{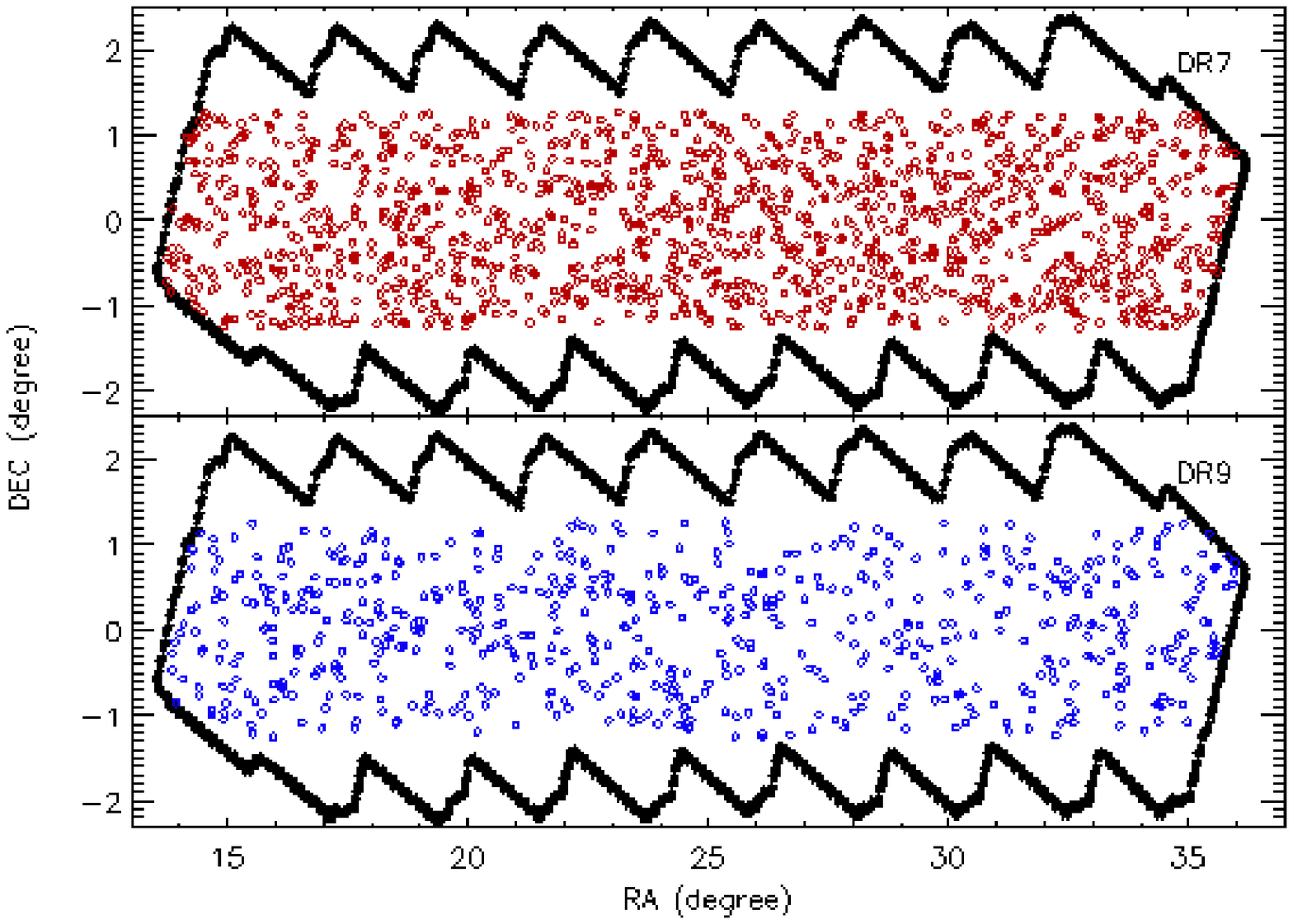}
\includegraphics[height=3.3in,width=3.35in]{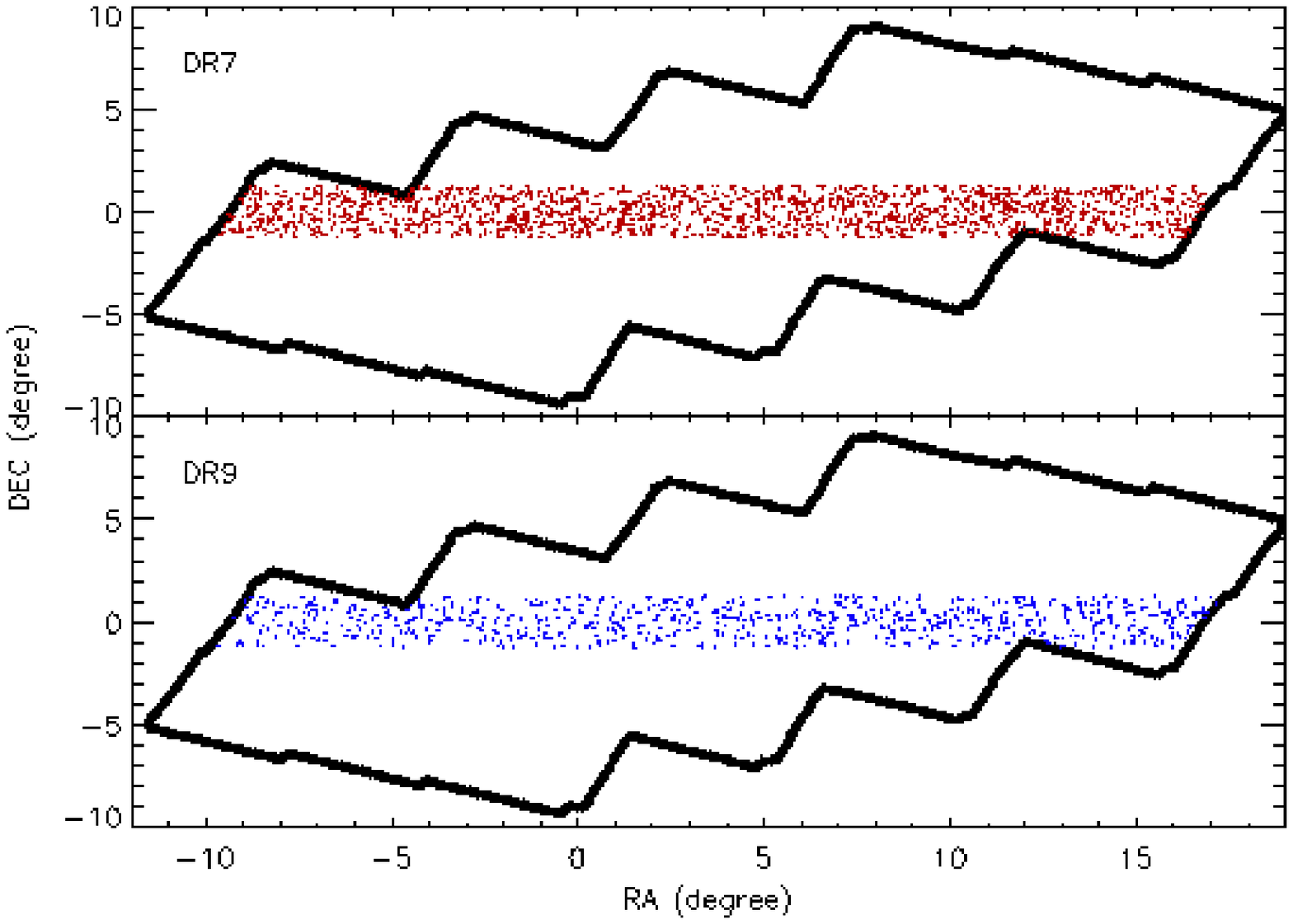}
\caption{Left: Sky distribution of the selected samples of the SDSS DR7 (top) and SDSS-III DR9 (bottom) quasars in the HeRS region. The black contours represent the HeRS footprint. Right: Sky distribution of the selected samples of the SDSS DR7 (top) and SDSS-III DR9 (bottom) quasars in the HeLMS region. The black contours represent the HeLMS footprint. For the DR9 QSOs, we only use the Stripe 82 region ($-1.25^{\circ} <$ Dec $< 1.25^{\circ}$) in both HeRS and HeLMS in our cross-correlation analysis. }
\label{fig:hers_helms_dr7_9}
\end{figure*}

In Fig.~\ref{fig:hers_helms_dr7_9}, we plot the sky distribution of the selected SDS DR7 and DR9 quasars in the region covered by the HeRS and HeLMS survey. Out of consideration for space, the distributions of the quasars in the Bo\"{o}tes, Lockman-SWIRE and {\it XMM}-LSS field are not shown here. For the DR7 QSO sample, we select all quasars at $0.15<z<2.5$ with $i$-band magnitude $<19.1$ (corrected for Galactic extinction) in the Bo\"{o}tes and Lockman-SWIRE region and $<19.9$ in the HeRS and HeLMS region. Based on visual inspection and lack of angular completeness masks, we assume that the angular completeness of the DR7 quasars is uniform. More discussion on this issue is presented in Section 3.1. For the DR9 QSO sample, we select all quasars at $2.2<z<3.5$ and ``uniform $>0$'' in Bo\"{o}tes, {\it XMM}-LSS, HeRS and HeLMS. We use DR9 quasar angular completeness masks from White et al. (2012) to model the completeness variations across the sky. 

We select the DR9 QSOs in HeRS and HeLMS to be within the Stripe 82 region ($-1.25^{\circ} <$ Dec $< 1.25^{\circ}$) and assume the angular completeness is uniform. In Table 1, we list the number of selected SDSS DR7 and DR9 QSOs found in the HeRS, HeLMS, Bo\"{o}tes, Lockman-SWIRE and {\it XMM}-LSS regions. Note that the number of DR9 QSOs in the HeRS and HeLMS Stripe 82 regions undergoes a final selection in Section 3.2.

\section{The cross-correlation between QSO and CIB}

\subsection{Cross-correlation estimator}

The cross-correlation of the two fields of interest $\rho_1$ (representing the quasar number density field) and $\rho_2$ (representing the CIB) is 
\begin{equation}
\xi=\left< \delta_1 \delta_2 \right>=\frac{1}{\overline{\rho}_1 \overline{\rho}_2} \left<(\rho_1 - \overline{\rho}_1)(\rho_2 - \overline{\rho}_2)\right>,
\end{equation}
where $\overline{\rho}_1$ is the mean number density of QSOs and $\overline{\rho}_2$ is the mean value of the CIB. The SPIRE maps, $\rho^\prime_2$, can be used to estimate $\rho_2$ but  have been mean-subtracted, i.e. the maps have a mean of zero $\overline{\rho^\prime_2}=0$.  The two quantities are thus related $\rho_2=\rho^\prime_2+C$, where $C$ is the unknown DC level of the map.  To proceed we assume a value for $C$ and add this to the map\footnote{We add a constant $C$ to our mean-subtracted maps so that the minimum value in the map is a positive number.} and will show that the choice of $C$ only affects the normalisation of our results. We define a re-normalised cross correlation $\xi^\prime$:\begin{eqnarray}
\xi^\prime=C \xi & = & \frac{C}{\overline{\rho}_1 (\overline{\rho}_2 + C)} \left< (\rho_1 - \overline{\rho}_1)\left(\rho^\prime_2 +C - (\overline{\rho^\prime_2 }+ C)\right)\right> \nonumber \\
  &= & \frac{1}{\overline{\rho}_1} \left< (\rho_1 - \overline{\rho}_1) (\rho^\prime_2) \right>
\end{eqnarray}
By construction, Eq. (3) gives the correlation between the QSO number density field and the FIR surface brightness field (with a normalisation constant). It is also equivalent to the stacked SPIRE map signal at the positions of the QSOs minus the mean signal of the map (which is equal to zero by design). This re-normalised cross-correlation, equivalent to stacking, is no longer dimensionless, but has the same unit as the map.

To estimate the two-point angular cross-correlation function $\xi$ between the quasar samples and the SPIRE maps, we  adapt the Landy \& Szalay (1993) auto-correlation estimator\footnote{The Landy \& Szalay  estimator is normally used to estimate correlation functions for point source catalogues. Here we use it to calculate the cross-correlation between a point source catalogue and an image.},
\begin{equation}
w^{\rm QS}(\theta) = \frac{D_1D_2(\theta) - D_1R_2(\theta) - R_1D_2(\theta) + R_1R_2(\theta)}{R_1R_2(\theta)}.
\end{equation}
Here $D_1$ represents the pair-counts in the real quasar sample, $D_2$ the CIB at a given wavelength, $R_1$ the simulated quasar sample with the same angular selection effects as the real quasar sample, and $R_2$ the simulated CIB with the same noise properties as the real SPIRE maps. $D_1D_2(\theta)$ represents the total emission in the CIB map at a separation of $\theta$ from  real quasars,  $D_1R_2(\theta)$ represents the total emission in the simulated CIB map at separation $\theta$ from all real quasars, $R_1D_2(\theta)$ represents the total emission in the CIB map at separation $\theta$ from all simulated quasars,  $R_1R_2(\theta)$ represents the total emission of the simulated CIB map at separation $\theta$ from all simulated quasars. Note that $D_1D_2$ is normalised by the total number of real quasars ($N_{\rm Q}$) times the total number of real SPIRE map pixels ($N_{\rm P}$), $D_1R_2$ is normalised by $N_{\rm Q}$ times the total number of simulated SPIRE map pixels (which is the same as $N_{\rm P}$), $R_1D_2$ is normalised by the total number of simulated quasars ($N_{S}$) times $N_{\rm P}$,  and $R_1R_2$ is normalised by $N_{S} \times N_{\rm P}$.

The CIB map and simulated CIB map are estimated from the mean-subtracted SPIRE maps and the simulated SPIRE maps by adding the same constant number $C$ to represent the mean CIB. As in Eq. (3) we then estimate a re-normalised cross-correlation signal. So, our actual estimator is\footnote{This is similar to stacking, shown in Appendix B. We have checked that the cross-correlation signal calculated using Eq. (5) is almost identical to the stacking result.}
\begin{eqnarray}
w^{\prime{\rm QS}}& = & C w^{\rm QS}(\theta) \nonumber\\
& = &C \frac{D_1D'_2(\theta)- D_1R'_2(\theta) - R_1D'_2(\theta) + R_1R'_2(\theta)}{R_1(R'_2(\theta)+C)}.
\end{eqnarray}
Here  $D'_2=D_2-C$ is the real SPIRE map, and $R'_2=R_2-C$ the simulated SPIRE map. We have checked that the cross-correlation result (calculated using Eq. (5)) as well as its error does not depend on the value of $C$ at all.  Although this is a reasonable choice of estimator, we have not investigated whether it is optimal.

The simulated SPIRE are constructed by randomising the pixels of the real map. It allows us to take into account of the effect of not only instrumental noise, but also confusion noise and cirrus. The random quasar samples are generated according to the angular completeness maps in the overlapping area between the SDSS DR7 and SDSS-III DR9 quasar sample and the \emph{Herschel} surveys. As mentioned before, we assume the angular completeness fraction is uniform for the DR7 quasars. So, we simply generate uniformly distribution points as the random DR7 quasar sample. Admittedly, the real angular completeness level for the DR7 quasars might vary across our fields. However, as what we are concerned with in this paper is the cross-correlation signal between the quasars and the CIB measured by {\emph Herschel}-SPIRE, the variation in the DR7 QSO completeness is not expected to bias our results as long as the completeness is uncorrelated with the noise properties of the SPIRE maps. For the DR9 QSOs in Bo\"{o}tes and {\it XMM}-LSS, we use the angular completeness masks used in White et al. (2012) and the MANGLE software (Swanson et al. 2008) to calculate the angular completeness of the survey, which is the fraction of quasar targets in a sector for which a spectrum was obtained. For the DR9 QSOs in the HeRS and HeLMS S82 region, we simply generate uniformly distributed points as the simulated QSO sample. The total number of simulated quasars $N_{S}$ is generally over 25 times larger than the total number of real quasars $N_{\rm Q}$.

We estimate errors on the two-point angular cross-correlation function by using 100 bootstrap samples of the quasar sample in each field. The bootstrap samples which have the same size as the real QSO sample are generated by sampling with replacement the real QSO sample. The dispersion in the signal between the bootstrap quasar samples and the SPIRE maps or the IRIS maps is taken as the error on the measured cross-correlation signal  (Norberg et al. 2009).

\begin{figure}
\includegraphics[height=2.7in,width=3.45in]{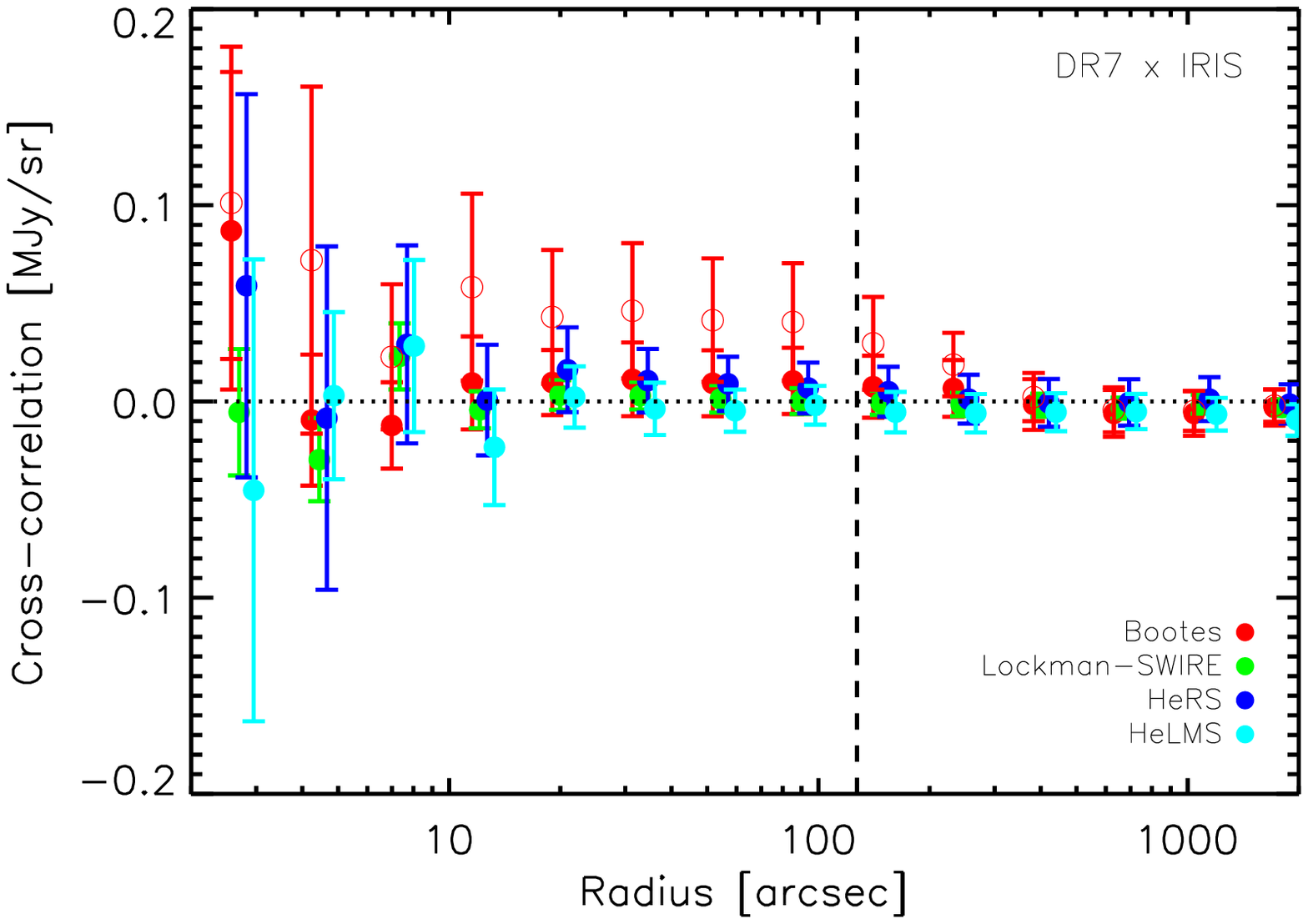}
\includegraphics[height=2.7in,width=3.45in]{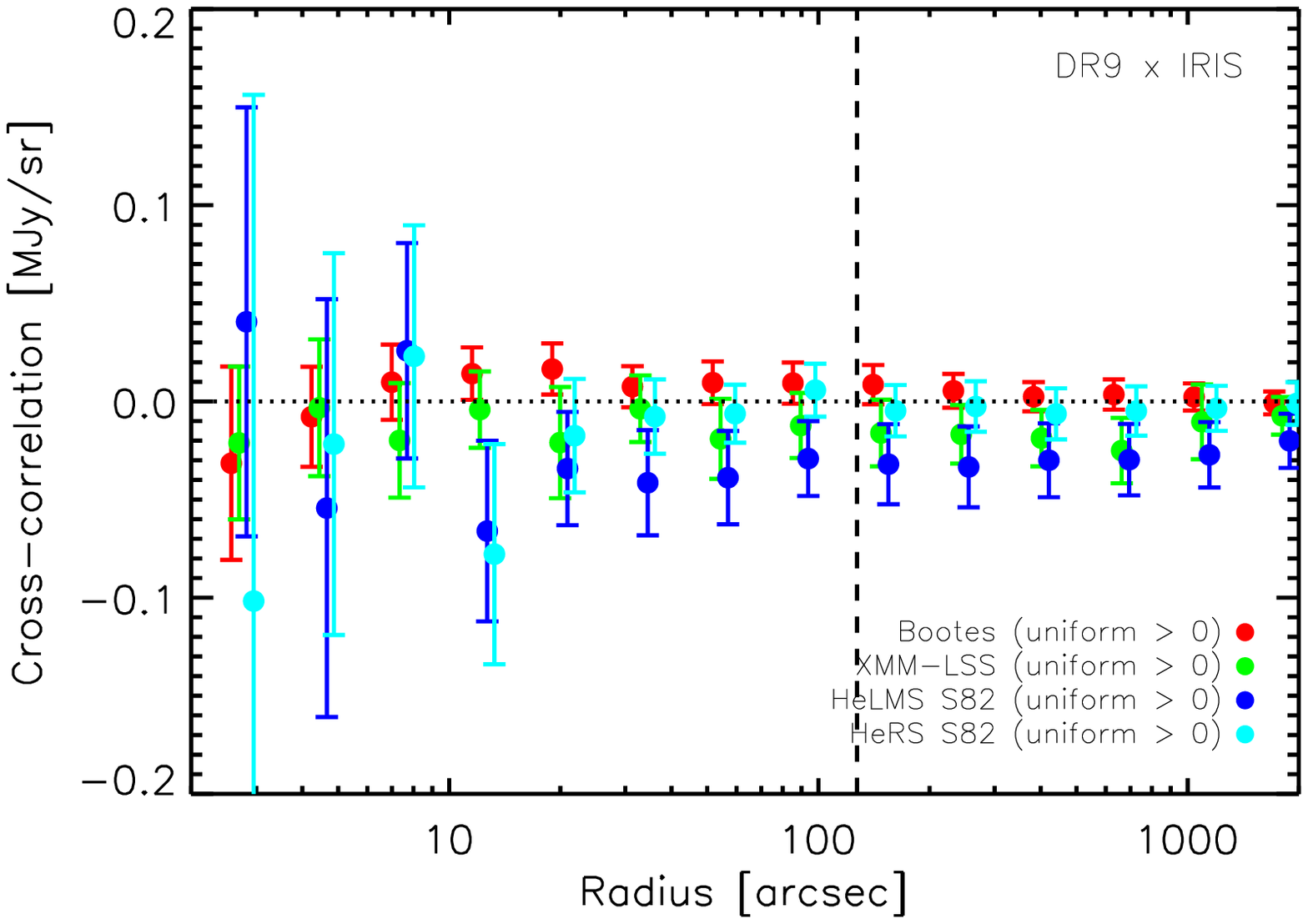}
\caption{Cross-correlation between the SDSS DR7 (top) and SDSS-III DR9 (bottom) QSOs and the 100\ $\micron$ maps. Different coloured symbols correspond to cross-correlation signal in different fields. One DR7 quasar in Bo\"{o}tes happens to lie on top of a very bright point source which biases that result, indicated by the empty red circles (therefore, it should be excluded from the sample). The filled red circles correspond to the cross-correlation between the new DR7 QSO sample in Bo\"{o}tes and the 100\ $\micron$ map. The horizontal  dotted line marks the zero signal expected if there is no correlation between quasars and cirrus. The vertical dashed line corresponds to the FWHM of the IRAS beam. Data points below the scale of the IRAS beam are strongly correlated. For the DR9 quasars, the cross-correlation with the 100\ $\micron$ maps in the HeLMS S82 region exhibits a systematic negative signal.  Data points from different fields are shifted slightly horizontally for visual clarity.}
\label{fig:qso_iris_all}
\end{figure}

\begin{figure}
\includegraphics[height=2.7in,width=3.45in]{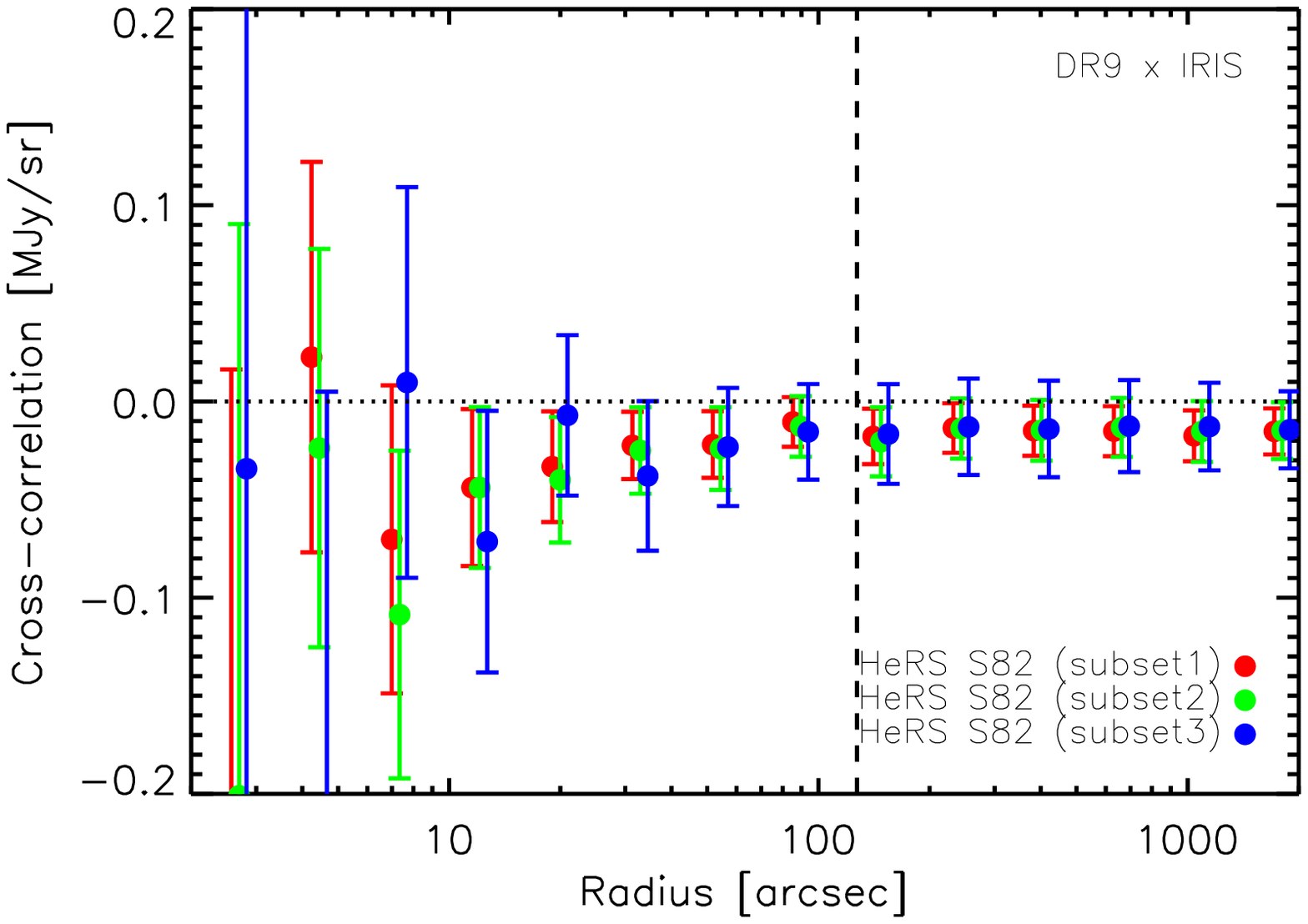}
\includegraphics[height=2.7in,width=3.45in]{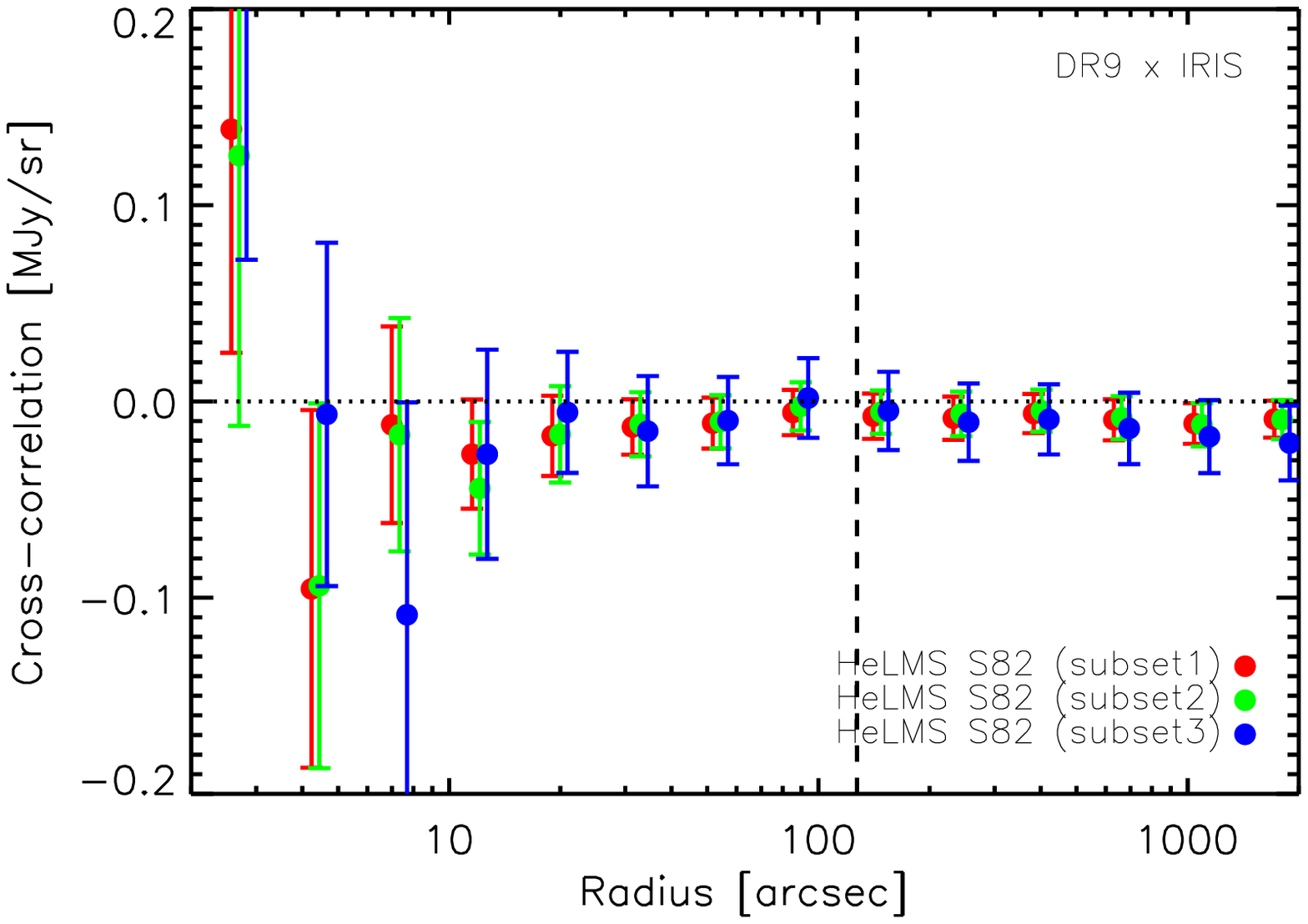}
\caption{Cross-correlation between DR9 QSOs and 100\ $\micron$ maps in the HeRS S82 (top) and HeLMS S82 (bottom) region. Different coloured symbols represent cross-correlation signal computed using DR9 quasars with different $i$-band apparent magnitude  cut (corrected for Galactic extinction) (red dots: $i$-band cut = 21.2; green dots: 21.0; blue dots: 20.0).  We have not used the ``uniform $>0$'' criterion in the samples plotted here. The anti-correlation in the HeLMS S82 region seen in Fig.~\ref{fig:qso_iris_all} has more or less disappeared. The horizontal dotted line corresponds to zero correlation. The vertical dashed line corresponds to the FWHM of the IRAS beam. Data points below the scale of the IRAS beam are strongly correlated. }
\label{fig:qso_iris_differentcuts_nonuniform}
\end{figure}

\begin{figure}
\includegraphics[height=2.7in,width=3.45in]{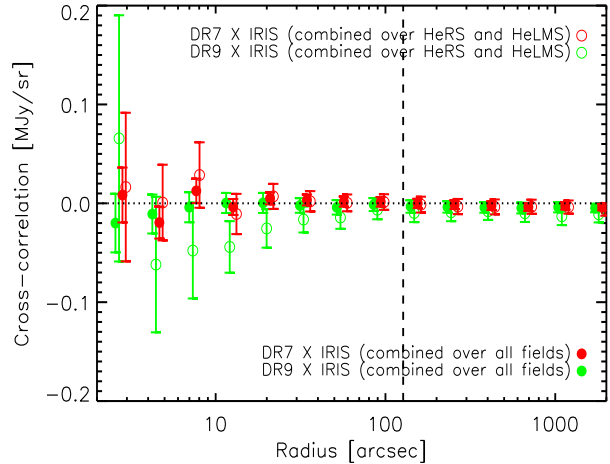}
\caption{The combined cross-correlation signal between the SDSS QSOs  (red symbols: DR7; green symbols: DR9)  and the 100\ $\micron$ maps, using inverse variance weighting. The filled circles represent the combined signal over all fields, while the open circles represent the combined results using only the HeRS and HeLMS regions. The horizontal dotted line corresponds to zero correlation. The vertical dashed line corresponds to the FWHM of the IRAS beam. Data points below the scale of the IRAS beam are strongly correlated. }
\label{fig:qso_iris_final}
\end{figure}

\subsection{Cross-correlation between quasars and IRIS 100\ $\micron$ maps}

We expect that there should be no intrinsic correlation between the quasars and Galactic cirrus. However, the observed quasar number density field could potentially (anti-)correlate with the intensity of the Galactic cirrus emission as the SDSS quasars are selected based on their optical flux and colour. In Fig.~\ref{fig:qso_iris_all}, we plot the cross-correlation signal between the quasar samples and the Galactic cirrus emission traced by the IRIS 100\ $\micron$ maps. The horizontal dotted line corresponds to no correlation. The vertical dashed line corresponds to the FWHM of the IRAS beam (Miville-Desch{\^e}nes et al. 2002). Data points below the scale of the IRAS beam are strongly correlated.  One quasar in Bo\"{o}tes happens to lie on top of a very bright point source in the IRIS map, which biases our result (the empty red circles in the top panel in Fig.~\ref{fig:qso_iris_all}). The filled red circles in the top panel in Fig.~\ref{fig:qso_iris_all} correspond to the cross-correlation between the DR7 QSOs, excluding the aforementioned quasar, and the 100\ $\micron$ map in Bo\"{o}tes. The reduced $\chi^2$ (the degrees of freedom DoF is 8), computed using data points above the scale corresponding to the FWHM of the IRAS beam, is 0.25, 0.24, 0.14 and 0.95 in the Bo\"{o}tes, Lockman-SWIRE, HeRS and HeLMS region, respectively. For the DR9 QSOs, the reduced $\chi^2$ (DoF = 8), computed using data points above the scale corresponding to the FWHM of the IRAS beam,  is 0.24, 0.92, 0.29 and 2.39 in the Bo\"{o}tes, {\it XMM}-LSS, HeRS S82 and HeLMS S82 region, respectively. The cross-correlation signal in the HeRS S82 region is consistent with zero, but is systematically negative in the HeLMS S82 region. 

The nominal $i$-band limit of the BOSS DR9 QSO sample is 21.8, which is deeper than the SDSS imaging limit of 21.3. So, in principle, it is possible that DR9 QSOs suffer more from incompleteness due to Galactic dust extinction.  We calculate the cross-correlation signal between ``uniform $>0$''  DR9 QSOs with different $i$-band magnitude cuts (corrected for Galactic extinction), namely 21.2, 21.0 and 20.0, and IRIS 100\ $\micron$ maps in the HeRS and HeLMS S82 region. The cross-correlation between different quasar sub-samples and the 100\ $\micron$ maps are consistent with each other. In the HeLMS S82 region, the cross-correlation signals between different sub-samples and the 100\ $\micron$ maps are still systematically below zero.  This test suggests that incompleteness due to Galactic dust extinction is not the likely cause for the anti-correlation seen in the HeLMS S82 region. 

In Fig.~\ref{fig:qso_iris_differentcuts_nonuniform}, we plot the cross-correlation signal between DR9 QSOs with different $i$-band magnitude cuts and IRIS 100\ $\micron$ maps in the HeRS and HeLMS S82 region. We have not used the ``uniform $>0$'' criterion in the samples plotted in Fig.~\ref{fig:qso_iris_differentcuts_nonuniform}. The anti-correlation in the HeLMS S82 region has now disappeared. The reduced $\chi2$ (DoF = 8), computed using data points above the scale corresponding to the FWHM of the IRAS beam, is 1.29, 0.84, 0.35 for DR9 QSOs with $i$-band cut of 21.2, 21.0 and 20.0 respectively in the HeRS region. The reduced $\chi2$ (DoF = 8), computed using data points above the scale corresponding to the FWHM of the IRAS beam, is 0.84, 0.68, 0.78 for DR9 QSOs with $i$-band cut of 21.2, 21.0 and 20.0 respectively in the HeLMS region. This makes sense as QSOs with ``uniform $>0$'' correspond to the XDQSO selection which is not the main selection method used in the S82 region. We finally select the DR9 QSOs with $i$-band cut of 21.0 in the HeRS and HeLMS S82 region, but not requiring ``uniform $>0$''.

In Fig.~\ref{fig:qso_iris_final}, we plot the combined cross-correlation signal between the quasars and the IRIS 100\ $\micron$ maps, using inverse variance weighting. The filled circles represent combined cross-correlation estimates over all fields and the empty circles represent combined cross-correlation estimates using only the HeRS and HeLMS fields. We can see clearly that the two types of combined estimates are consistent with each other. The cross-correlation between either DR7 or DR9 QSOs and the 100\ $\micron$ maps are consistent with zero. As we have discussed in Section 2.1, the dominant signal in the 100\ $\micron$ maps are due to Galactic cirrus emission (especially in the HeRS and HeLMS regions). We conclude that the correlations between the final selected QSO samples and Galactic cirrus are consistent with zero.

\subsection{Cross-correlation between quasars and SPIRE 250, 350 and 500\ $\micron$ maps}

\begin{figure*}
\includegraphics[height=2.7in,width=3.45in]{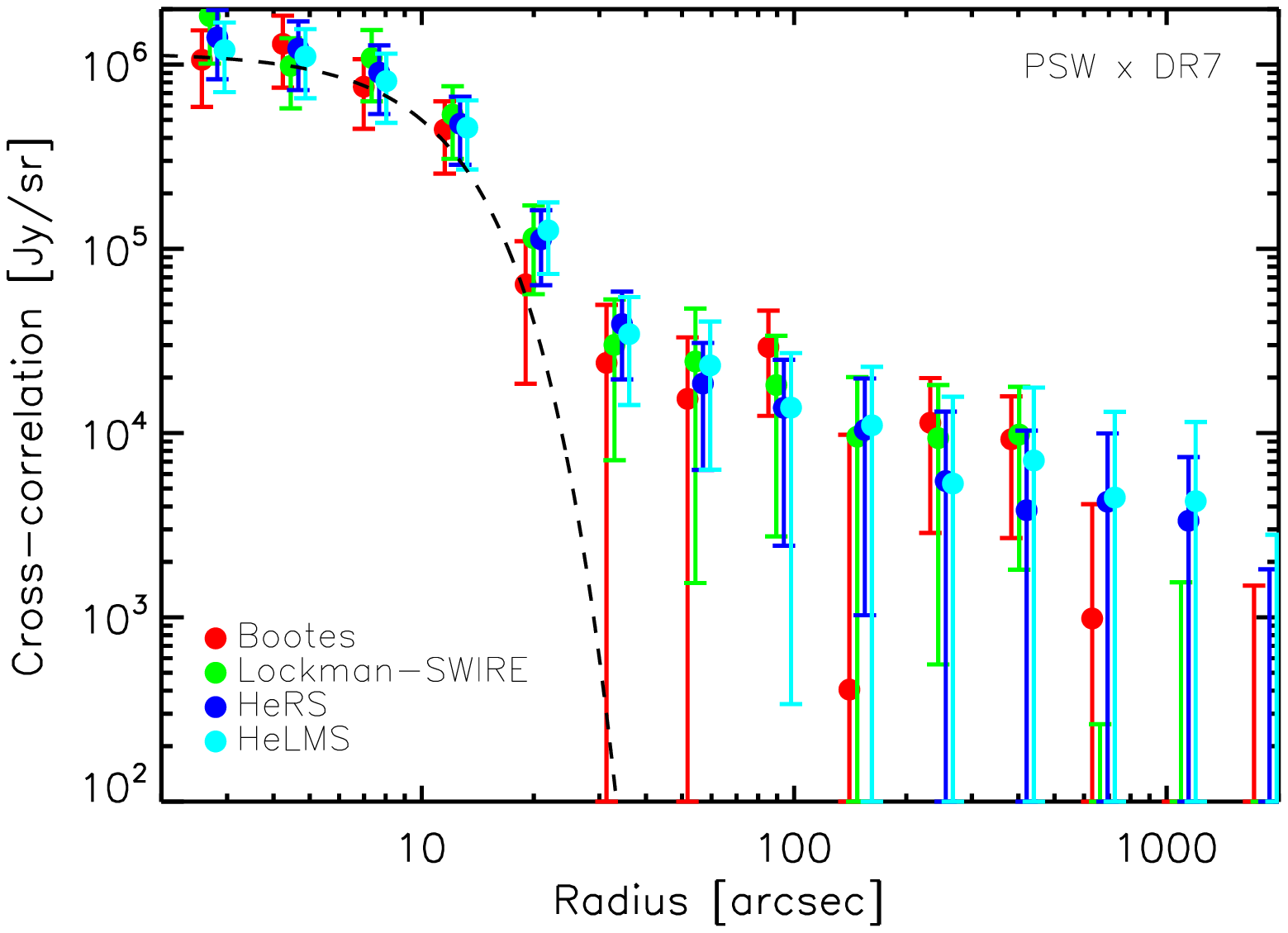}
\includegraphics[height=2.7in,width=3.45in]{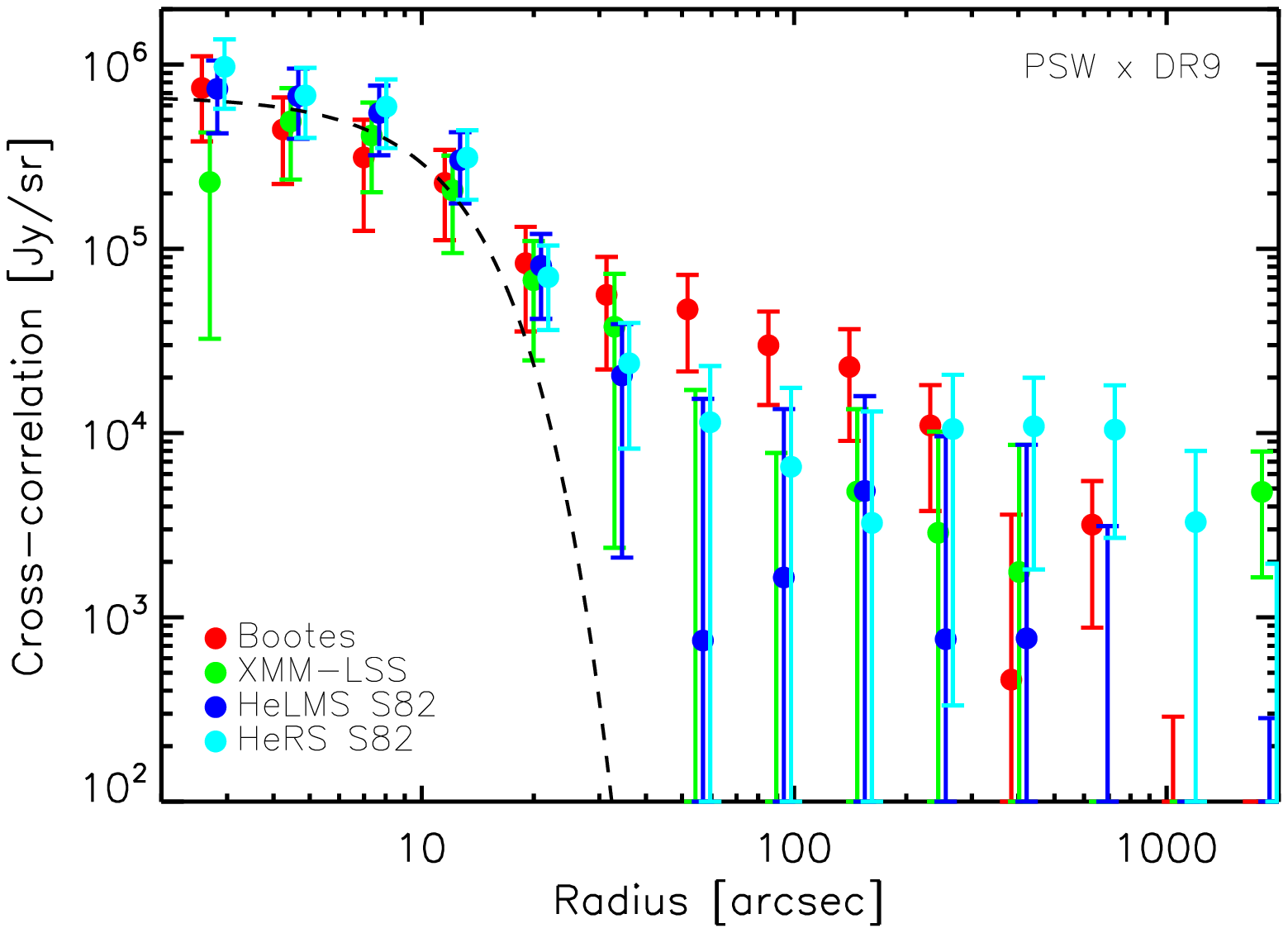}
\includegraphics[height=2.7in,width=3.45in]{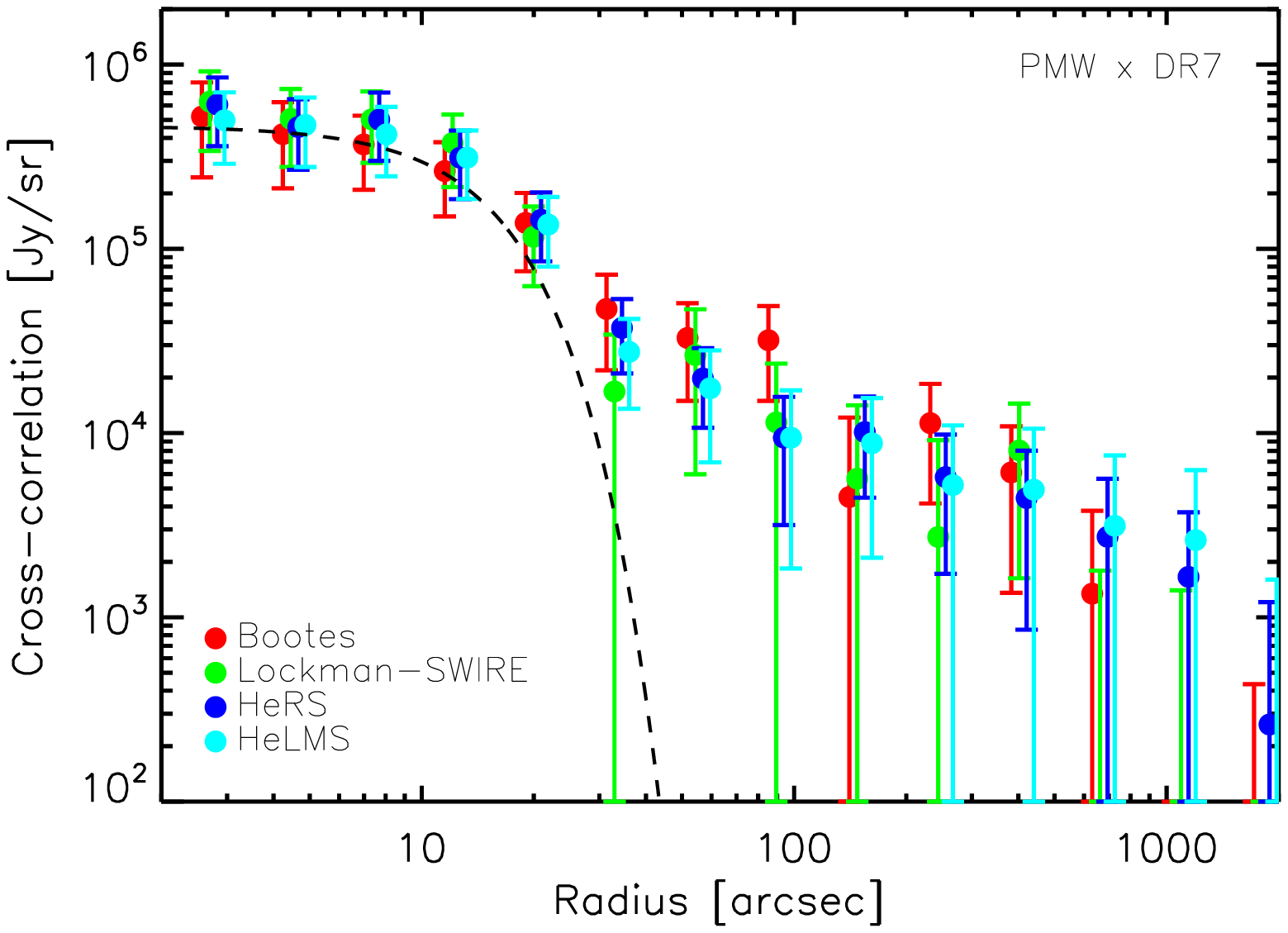}
\includegraphics[height=2.7in,width=3.45in]{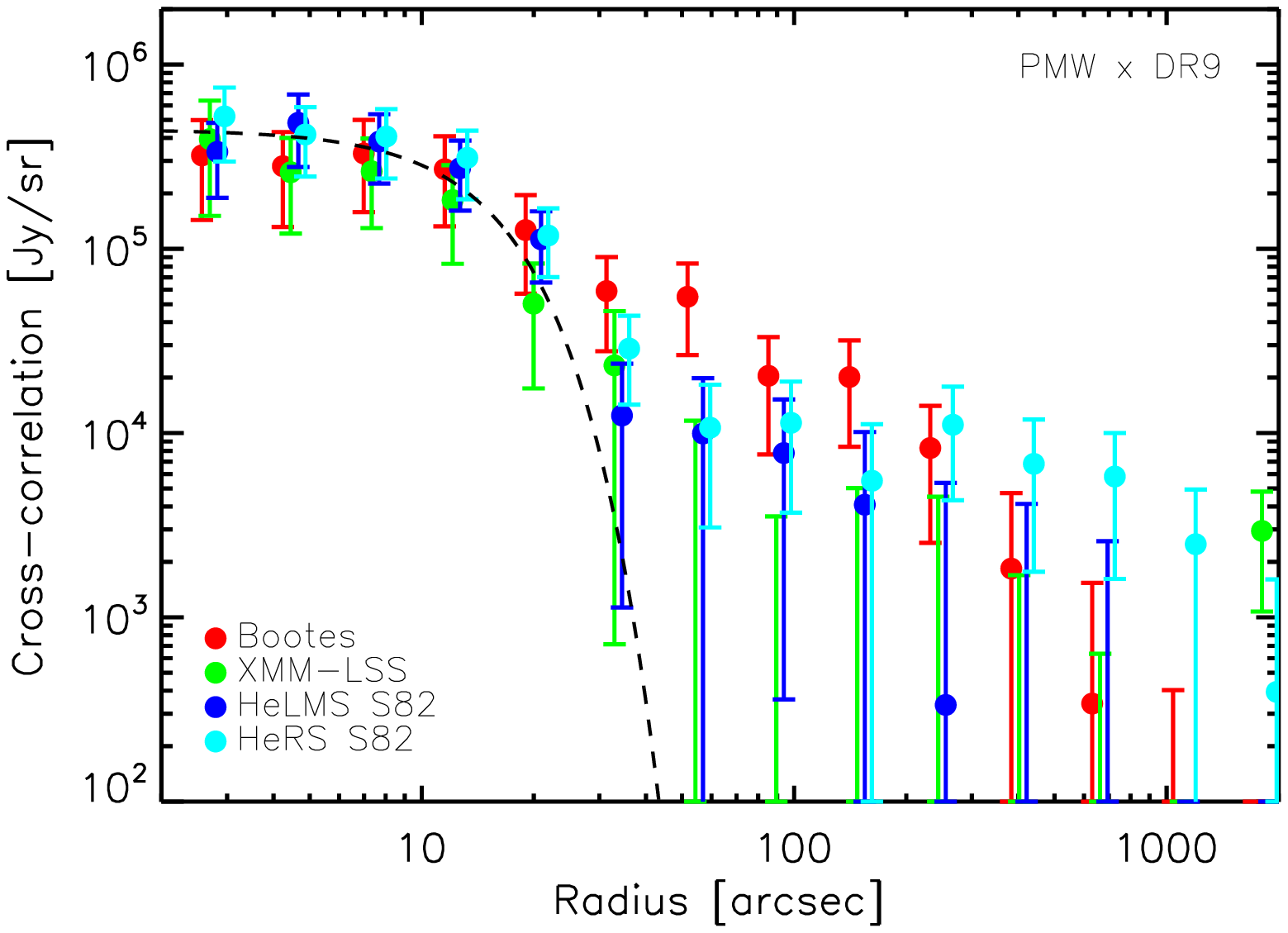}
\includegraphics[height=2.7in,width=3.45in]{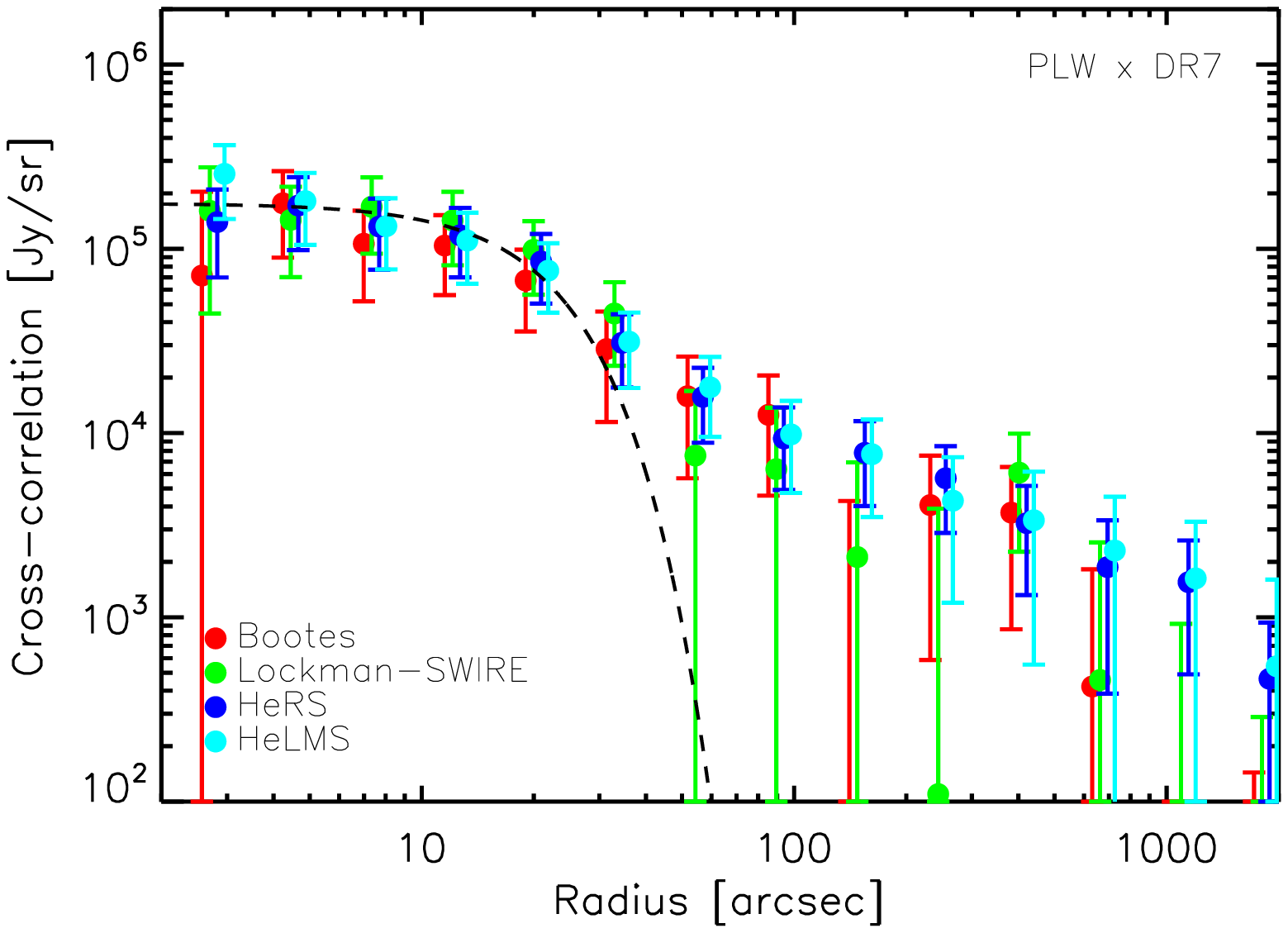}
\includegraphics[height=2.7in,width=3.45in]{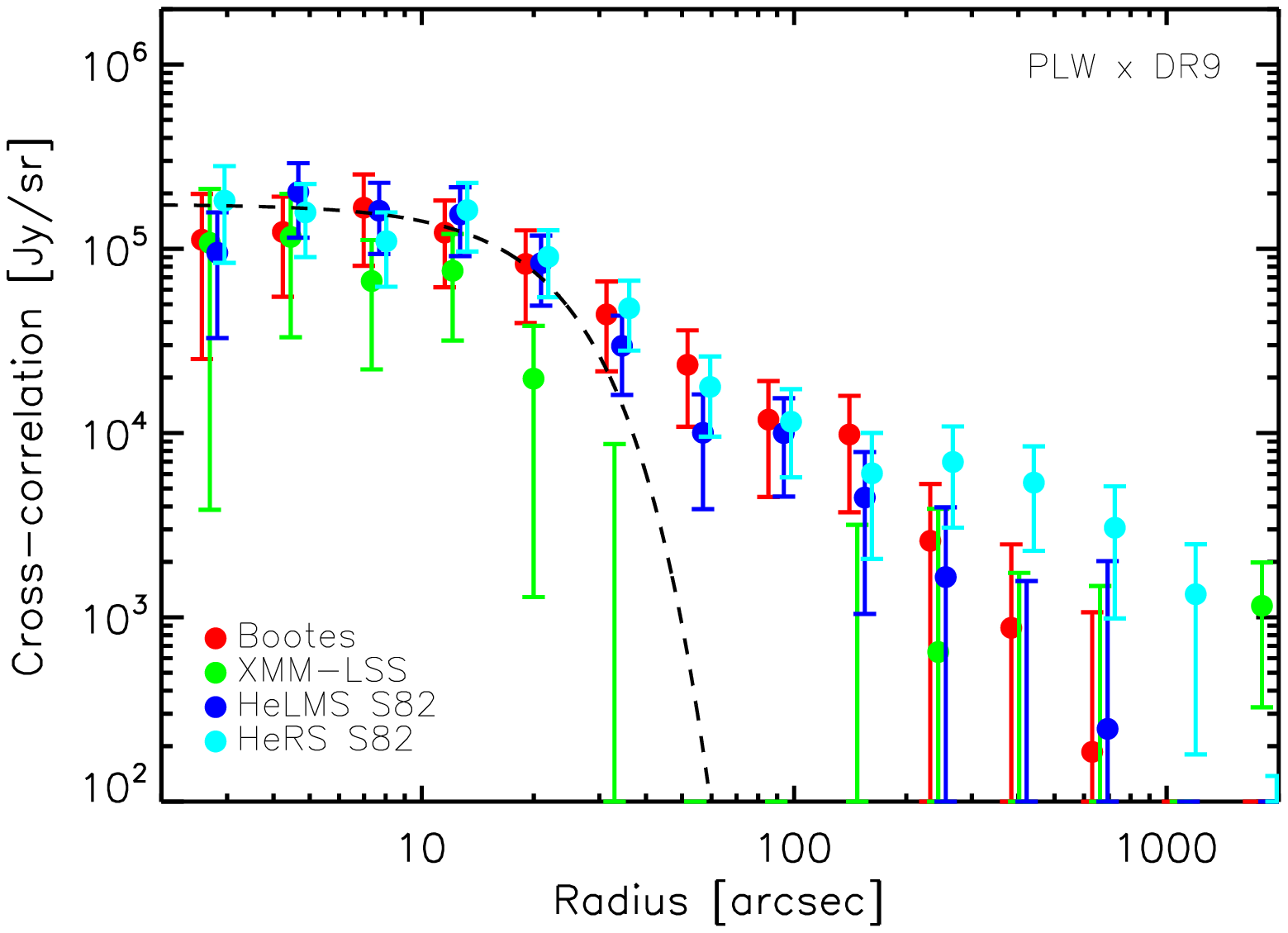}
\caption{The cross-correlation signal between QSOs (left column: SDSS DR7 QSOs in the redshift range $z=[0.15, 2.5]$; right column: SDSS-III DR9 QSOs in the redshift range $z=[2.2, 3.5]$) and the CIB at 250 (top), 350 (middle) and 500 (bottom) $\micron$. Different coloured symbols correspond to cross-correlation signal in different fields. Data points from different fields are shifted slightly horizontally for visual clarity. The dashed line in each panel corresponds to the SPIRE PSF in the corresponding waveband. No fitting has been carried out here, so the amplitude of the PSF is approximate. The cross-correlation signal in different fields generally give consistent result. On small angular scales, the cross-correlation signal is dominated by the mean sub-mm emission of the quasars themselves. On larger scales, an excess signal beyond the SPIRE PSF is seen in almost all cases, indicating that there are DSFGs correlated with the quasars. These correlated DSFGs may include satellite galaxies in the same halo as the quasar and galaxies in separate halos correlated with the quasar-hosting halos.}
\label{fig:corr_signal}
\end{figure*}

In the left column of Fig.~\ref{fig:corr_signal}, we plot the cross-correlation signal between the SDSS DR7 QSOs (in the redshift range $z=[0.15, 2.5]$) and the SPIRE maps at 250, 350 and 500\ $\micron$ in HeRS, HeLMS, Bo\"{o}tes and Lockman-SWIRE. The cross-correlation signal in  four different fields are consistent with each other. On small scales, the cross-correlation signal is dominated by the mean sub-mm emission of the quasars themselves, as indicated by the SPIRE point spread function (PSF) in the corresponding waveband. No fitting to the measurement is carried out at this stage, so the amplitude of the PSF is only approximate. On larger scales, significant excess signal beyond the SPIRE PSF is seen in all three wavebands in all fields, indicating that there is a population of DSFGs correlated with the quasar population. They could be mainly satellite galaxies residing in the same dark matter halo as the host galaxy of the quasar and galaxies in other halos which are correlated with the halo hosting the quasar.


In the right column Fig.~\ref{fig:corr_signal}, we plot the cross-correlation signal between the SDSS-III DR9 QSOs (in the redshift range $z=[2.2, 3.5]$) and the SPIRE maps at 250, 350 and 500\ $\micron$ in HeRS, HeLMS, Bo\"{o}tes and {\it XMM}-LSS.  Again, on small scales, the cross-correlation signal is dominated by the mean sub-mm emission of the quasars themselves, as indicated by the SPIRE beams. Compared to the DR7 quasar sample, the amplitude of the SPIRE emission of the DR9 quasar sample is slightly smaller in all three bands. On larger scales, excess signal beyond the PSF is seen in all fields at all wavelengths apart from the {\it XMM}-LSS field at 350 and 500\ $\micron$. The dispersion in the cross-correlation signal between different fields is also larger for the DR9 QSOs than the DR7 QSOs.  It could be due to the surface density of the DR9 quasars which is around a factor of two lower than the surface density of the DR7 quasars (except in Bo\"{o}tes).

\begin{figure}
\includegraphics[height=2.65in,width=3.45in]{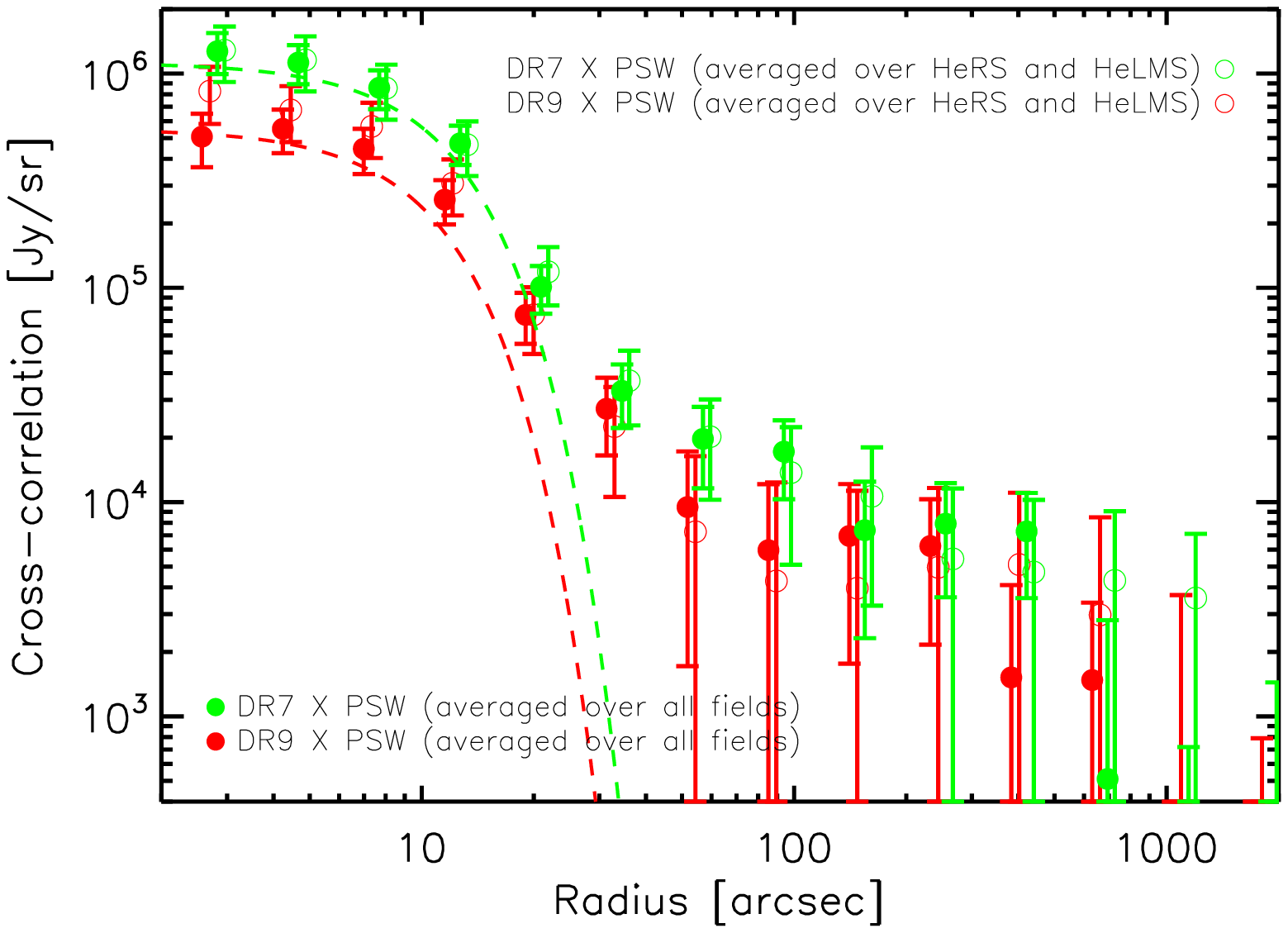}
\includegraphics[height=2.65in,width=3.45in]{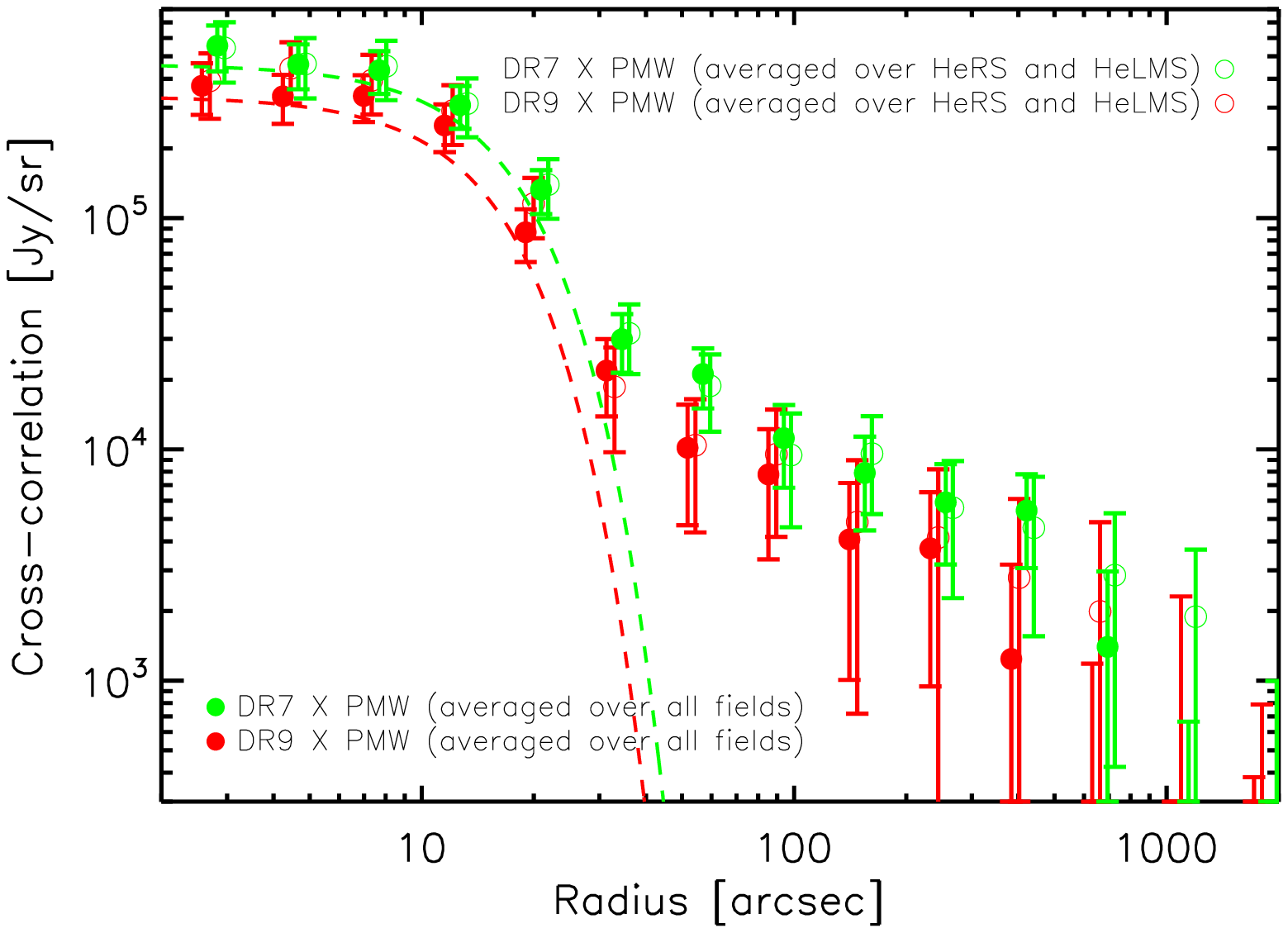}
\includegraphics[height=2.65in,width=3.45in]{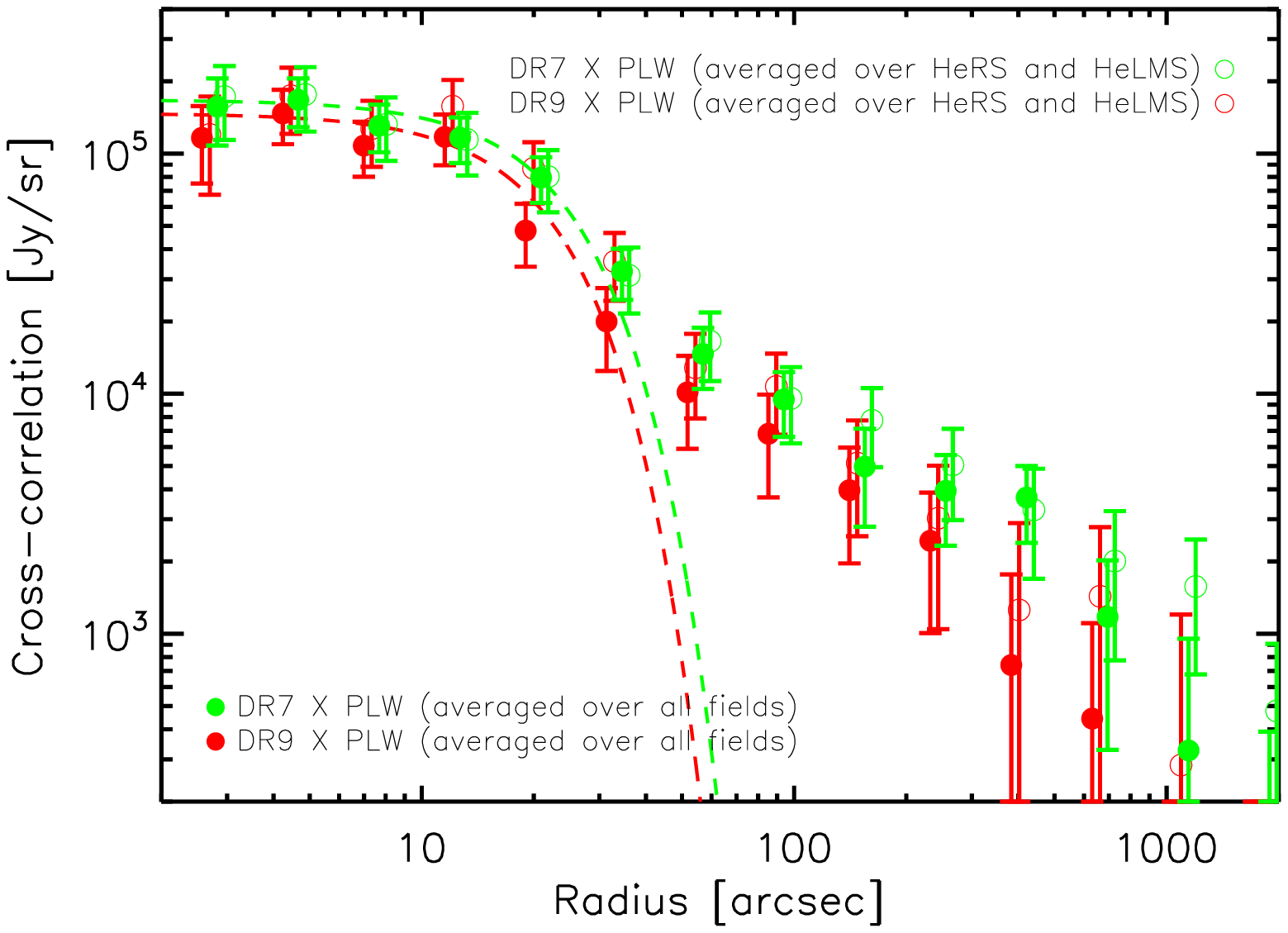}
\caption{The combined cross-correlation signal between QSOs (green symbols: DR7; red symbols: DR7) and the SPIRE maps at 250, 350 and 500 $\micron$ (filled circles: combined signal averaged over all fields; empty circles: combined signal averaged over HeRS and HeLMS). The two different types of combined signal agree well with each other in all cases. The dashed lines in each panel corresponds to the SPIRE PSF in the corresponding waveband. No fitting has been carried out here, so the amplitude of the PSF is approximate. Data points from different wavebands are shifted slightly horizontally for visual clarity.}
\label{fig:corr_signal_final_sbands}
\end{figure}

\begin{figure}
\includegraphics[height=2.7in,width=3.45in]{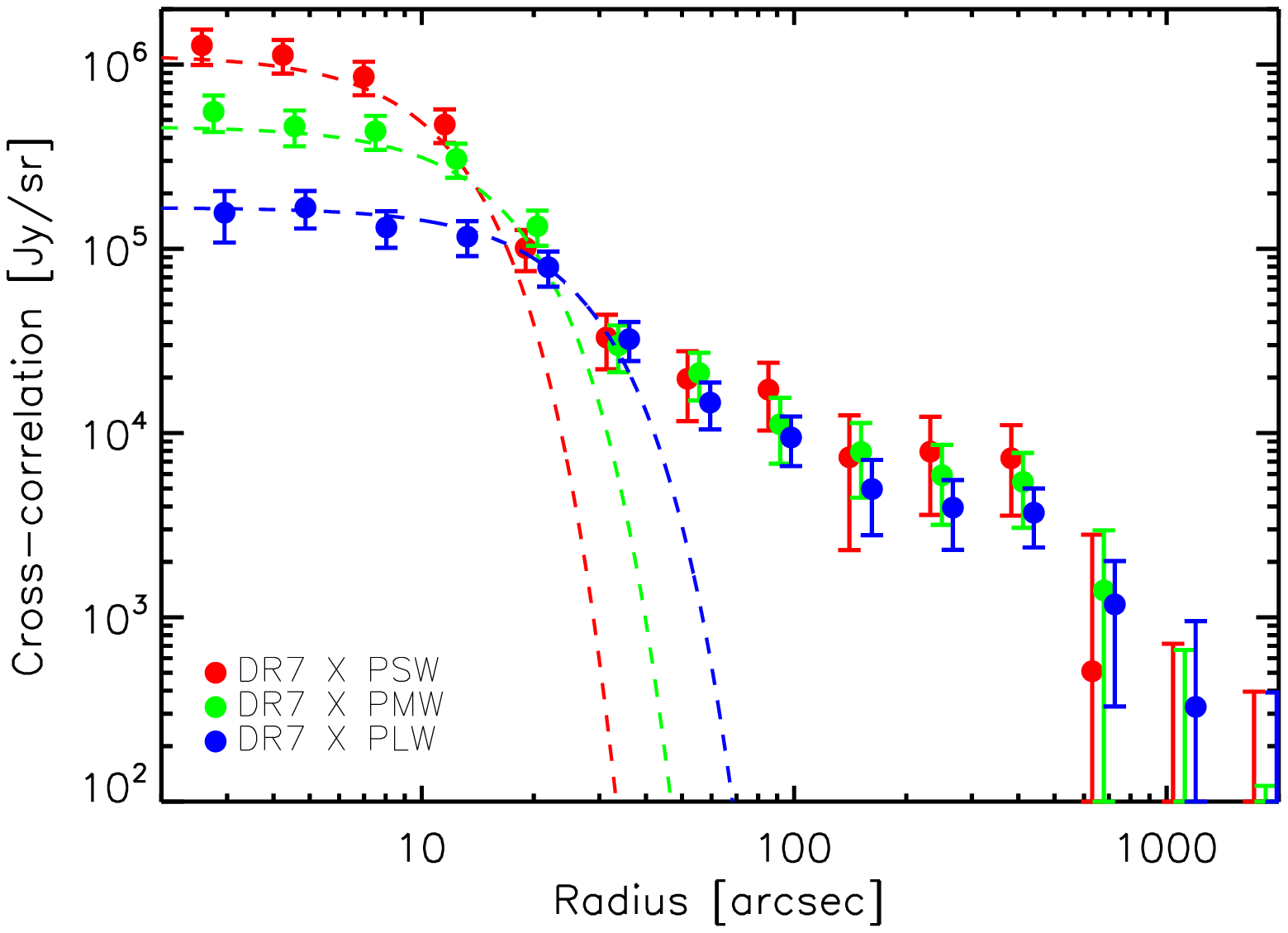}
\includegraphics[height=2.7in,width=3.45in]{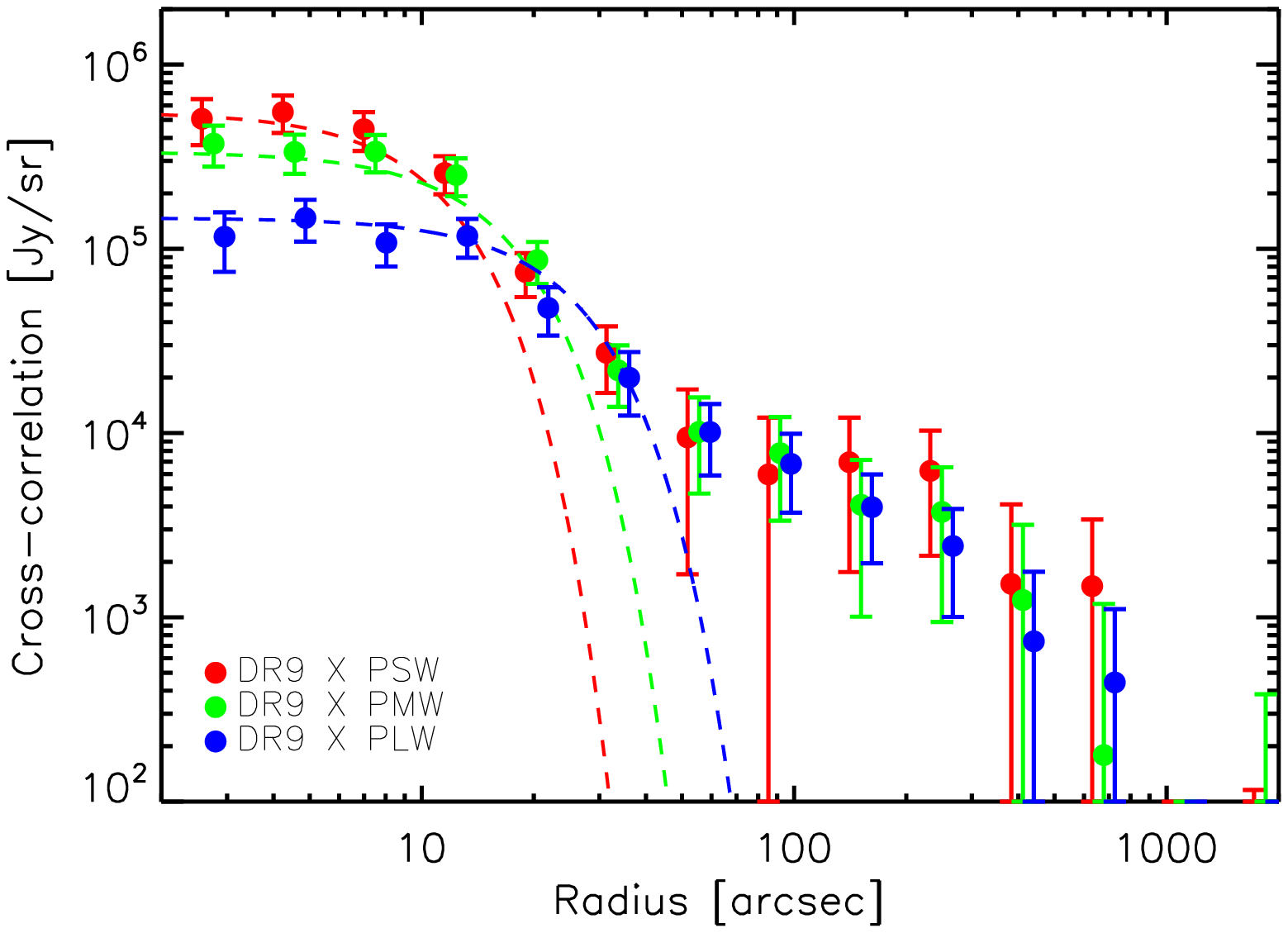}
\caption{The combined cross-correlation signal between QSOs (top: DR7; bottom: DR9) and the SPIRE maps at 250, 350 and 500 $\micron$, averaged over all fields. Different coloured symbols correspond to cross-correlation signal between QSOs and SPIRE maps at different wavelengths. The dashed lines in each panel corresponds to the SPIRE PSF in the corresponding waveband (red: 250\ $\micron$; green: 250\ $\micron$; blue: 500\ $\micron$). No fitting has been carried out here, so the amplitude of the PSF is approximate. Data points from different wavebands are shifted slightly horizontally for visual clarity.}
\label{fig:corr_signal_final}
\end{figure}

In Fig.~\ref{fig:corr_signal_final_sbands}, we plot the combined cross-correlation signal between the QSOs and the CIB at 250, 350 and 500\ $\micron$, using inverse variance weighting. The filled circles represent combined cross-correlation estimates over all fields and the empty circles represent combined cross-correlation estimates using only the HeRS and HeLMS fields. The two different types of combined cross-correlation signal are consistent with each other and exhibit significant excess signal beyond the SPIRE PSF at all wavelengths. In Fig.~\ref{fig:corr_signal_final}, we compare the combined cross-correlation signal (using all fields) between the SDSS QSOs and the SPIRE maps at all three wavelengths. For both DR7 and DR9 quasars, the sub-mm emission of the quasars themselves decreases by almost an order magnitude from 250 to 500\ $\micron$, while the change in the correlated sub-mm emission is much smaller from 250 to 500\ $\micron$. This is partly to do with the broadening of the SPIRE beam from 250 to 500\ $\micron$ which affects point sources (i.e. the sub-mm emission of the quasars themselves) more than extended structures (i.e. the correlated signal from DSFGs). Another possibility is that the QSOs have a higher effective dust temperature than that of the correlated DSFGs. 


\section{Halo model of the co-evolution of black hole growth and star formation}

\subsection{Halo model of the CIB}

The halo model of the CIB anisotropies presented in this paper is identical to the model presented in Viero et al. (2013) and Hajian et al. (in prep), which is in turned developed based on {Shang et al. (2013)}. 
For the sake of completeness, we outline the main elements of the CIB model here. For more details, please refer to Shang et al. (2013), Viero et al. (2013) and Hajian et al. (in prep).

The power spectra of the CIB sourced by infrared galaxies is composed of a Poisson noise (or shot noise) term and a clustered term,
\begin{equation}
P(k_{\theta})_{\nu_1 \nu2}^{\rm CIB} = P(k_{\theta})_{\nu_1 \nu2}^{\rm CIB-Poisson} + P(k_{\theta})_{\nu_1 \nu2}^{\rm CIB-clust}. 
\end{equation}
The Poisson noise term in the CIB power spectrum, independent of angular scale, can be calculated from the multi-dimensional number counts of the infrared sources,
\begin{equation}
P(k_{\theta})_{\nu_1 \nu2}^{\rm CIB-Poisson} = \int_0^{S_{\rm cut}} S_{\nu_1} S_{\nu_2} \frac{d^2 N}{d S_{\nu_1} d S_{\nu_2}} d S_{\nu_1} d S_{\nu_2}.
\end{equation}
The clustered term in the angular power spectrum of the CIB is the flux averaged projection of the three-dimensional (3D) spatial power spectrum,
\begin{equation}
P(k_{\theta})_{\nu_1 \nu_2}^{\rm CIB-clust} = \int \frac{dz}{dV_c/dz} P_{\nu_1 \nu_2} (k, z)^{\rm CIB-clust} \frac{dS_{\nu_1}}{dz}  \frac{dS_{\nu_2}}{dz}.
\end{equation}
Here $dV_c$ is the comoving volume element, and $\frac{dS_{\nu}}{dz}$ is the redshift distribution of the integrated flux at a given frequency which can be calculated as
\begin{equation}
\frac{dS_{\nu}}{dz} = \int_0^{S_{\rm cut}} S_{\nu} \frac{d^2N}{dS_{\nu} dz} dS_{\nu}.
\end{equation}
The 3D spatial power spectrum of CIB sources can be decomposed into a 1-halo and 2-halo term which describes the small-scale clustering due to pairs of galaxies in the same dark matter halo and the large-scale clustering due to pairs of galaxies in different halos respectively,
\begin{equation}
P(k,z)_{\nu_1 \nu_2}^{\rm CIB-clust} = P(k,z)_{\nu_1 \nu2}^{\rm CIB-clust:1h} + P(k,z)_{\nu_1 \nu_2}^{\rm CIB-clust:2h}.
\end{equation}
The 3D non-linear, 1-halo term of the CIB clustered term can be calculated as
\begin{eqnarray}
&&P(k,z)_{\nu_1 \nu2}^{\rm CIB-clust:1h} = \frac{1}{\bar{j}_{\nu_1} \bar{j}_{\nu_2}}\int dM \frac{dN}{dM} (z)\\ \nonumber
 && \times [f_{\nu_1}^{\rm cen}(M,z) f_{\nu_2}^{\rm sat}(M,z) u_{\rm gal}(k,z|M)\\ \nonumber
&&+f_{\nu_2}^{\rm cen}(M,z) f_{\nu_1}^{\rm sat}(M,z) u_{\rm gal}(k,z|M) \\ \nonumber
&&+ f_{\nu_1}^{\rm sat}(M,z) f_{\nu_2}^{\rm sat}(M,z) u_{\rm gal}^2(k,z|M)].\\
\end{eqnarray}
Here $f_{\nu}^{\rm cen} (M, z)$ and $f_{\nu}^{\rm sat} (M, z)$ are the luminosity-weighted number of central and satellite galaxies as a function of redshift and halo mass, $\bar{j}_{\nu}(z)$ is the mean comoving specific emission coefficient as a function of redshift, $\frac{dN}{dM}(z)$ is the redshift-dependent halo mass function (taken from Tinker et al. 2008), $u_{\rm gal}(k,z|M)$ is the normalised Fourier transform of the galaxy density distribution within a halo (assumed to follow the NFW profile in this paper).
Here luminosity-weighted number of central and satellite galaxies can be derived as
\begin{equation}
f_{\nu}^{\rm cen} (M, z) = N^{\rm cen} \frac{L_{(1+z)\nu}^{\rm cen}(M, z)}{4\pi}
\end{equation}
and
\begin{equation}
f_{\nu}^{\rm sat} (M, z) = \int dm \frac{dn}{dm} (M, z) \frac{L_{(1+z)\nu}^{\rm sat}(m, z)}{4\pi},
\end{equation}
where $m$ is the subhalo mass at the time of accretion and $\frac{dn}{dm} (M, z)$ is the sub-halo mass function (taken from Tinker \& Wetzel 2010). Therefore, the mean comoving specific emission coefficient is
\begin{equation}
\bar{j}_{\nu} = \int dM \frac{dN}{dM} (z) \frac{1}{4\pi} [f_{\nu}^{\rm cen} (M, z) + f_{\nu}^{\rm sat} (M, z)].
\end{equation}
We assume a log-normal relation between halo mass and infrared luminosity,
\begin{equation}
\Sigma(M) = L_0 M \frac{1}{\sqrt{2\pi\sigma^2_{L/M}}}\exp{\left[-\frac{(\log M - \log M_{\rm peak})^2}{2\sigma^2_{\rm L/M}}\right]},
\end{equation}
where $M_{\rm peak}$ and $\sigma_{L/M}$ is peak halo mass scale and $1\sigma$ range of the specific infrared luminosity per unit mass, and $L_0$ is the overall infrared luminosity to halo mass normalisation factor. A lower limit on the halo mass $M_{\rm min}$ is applied to the halo mass - infrared luminosity relation.
To describe the spectral energy distribution (SED) of the infrared galaxy population, we use a single modified blackbody with two parameters, i.e. dust temperature $T_{\rm dust}$ and emissivity $\beta$.
In addition, the infrared luminosity is assumed to evolve with redshift as $(1+z)^{\eta}$ over the range $0<z<2$ followed by a plateau at $z>2$. The 2-halo term of the CIB clustering power spectrum can be calculated as
\begin{equation}
P(k,z)_{\nu_1 \nu_2}^{\rm CIB-clust:2h} = \frac{1}{\bar{j}_{\nu_1} \bar{j}_{\nu_2}} P_{\rm lin}(k,z) D_{\nu_1}(k,z) D_{\nu_2} (k,z),
\end{equation}
where $P_{\rm lin}(k,z)$ is the linear dark matter power spectrum and 
\begin{eqnarray}
D_{\nu_1}(k, z) &= &\int dM \frac{dN}{dM} b(M,z) u_{\rm gal}(k,z|M) \\ \nonumber
&&\times [f_{\nu_1}^{\rm cen}(M,z) + f_{\nu_2}^{\rm sat}(M,z)].
\end{eqnarray}
Here $b(M,z)$ is the linear large-scale bias. We use the prescription for the halo bias from Tinker, Wechsler \& Zheng (2010) in this paper.

Due to the limited constraining power of the measured cross-correlation signal between the QSOs and the SPIRE maps (our measurements are quite noisy especially on large scales), in this paper we have used the same parameters of the CIB halo model as in Viero et al. (2013) which fits the auto- and cross-correlation power spectra of the SPIRE bands. The exact values of the parameters used in the CIB model are minimum halo mass $M_{\rm min}=10^{10.1} M_{\odot}$, dust temperature $T_{\rm dust} = 24.2$ K, peak halo mass $M_{\rm peak} = 10^{12.1} M_{\odot}$, $\sigma^2_{L/M}=0.38$, infrared luminosity to halo mass normalisation $L_0=10^{-1.71}L_{\odot}/M_{\odot}$, dust emissivity $\beta=1.45$ and luminosity evolution $\eta=2.19$.

\subsection{Halo model of the quasar population}

Many studies have used the clustering measurements of quasars to constrain their halo occupation distributions. Generally, quasars are found to inhabit dark matter halos of masses around a few times $10^{12} M_{\odot}$ (Croom et al. 2005; Da Angela et al. 2008; Shen et al. 2009; Ross et al. 2009; White et al. 2012; Shen et al. 2012). The clustering strength of quasars seems to depend very weakly on QSO luminosity or redshift. It indicates that there is a  poor correlation between halo mass and the instantaneous quasar luminosity. Richardson et al. (2012) found tentative evidence for increasing halo mass scale for quasars with increasing redshift. They found that at $z\sim1.4$ the median halo mass hosting central quasars is around $4.1^{+0.3}_{-0.4}\times10^{12}M_{\odot}$ while at $z\sim3.2$ the median halo mass increases to $14.1^{+5.8}_{-6.9}\times10^{12}M_{\odot}$. Most quasars are central galaxies according HOD analysis of quasar clustering statistics. The satellite fraction of quasars is generally found to be of the order a few percent at most, although the exact satellite fraction estimated from different studies vary significantly from each other. For example, Shen et al. (2013) using a cross-correlation study between the DR7 quasars and DR10 BOSS galaxies estimated the satellite fraction to be  $0.068^{+0.034}_{-0.023}$. A similar halo model applied to the auto-correlation of the DR7 quasars combined with a binary quasar sample (Hennawi et al. 2006) in Richardson et al. (2012) inferred the satellite fraction to be $(7.4\pm1.3)\times 10^{-4}$. 

In this paper, we use a simple halo model for the quasars, widely used  in the literature. The quasar mean halo occupation number is composed of central and satellite components,
\begin{equation}
\left< N_q(M)\right> = \left< N_q^{\rm cen}(M)\right>  + \left< N_q^{\rm sat}(M)\right> 
\end{equation}
A softened step function is assumed to describe the central component,
\begin{equation}
N_q^{\rm{cen}}(M) = \frac{1}{2} \left[1 + \rm{erf}\left(\frac{log M - log M_{\rm min}}{\sigma_{logM}}\right)\right]
\end{equation}
and a rolling-off power law for the satellite component, 
\begin{equation}
N_q^{\rm{sat}}(M) = (M/M_1)^{\alpha} \exp{(-M_{\rm cut} / M)}.
\end{equation}
There are five free parameters in the quasar HOD: $M_{\rm min}$ is the characteristic halo mass scale at which on average half of the dark matter halos host one quasar as the central galaxy; $\sigma_{\log M}$ is the width of the softened step function; $M_1$ is the approximate halo mass scale where on average halos host one satellite quasar; $\alpha$ is the power-law index; $M_{\rm cut}$ is the halo mass scale below which the number of the satellite quasars decreases exponentially.

In principle, we should also multiply a free parameter to the quasar HOD $N_q(M)$ to describe the duty cycle of quasars, i.e. only a certain fraction ($<1$) of the halos actually host quasars. However, if we assume that quasar duty cycle is independent of halo mass, then it will cancel the duty cycle parameter which will appear in the QSO number density $\bar{n}$ in Eq. (21) and (22). So, the quasar HOD equations (Eq. 18 -- 20) should be interpreted as the number of quasars in halos of mass $M$ which host quasars (not all halos of mass $M$). 

\begin{figure*}
\includegraphics[height=4.in,width=3.45in]{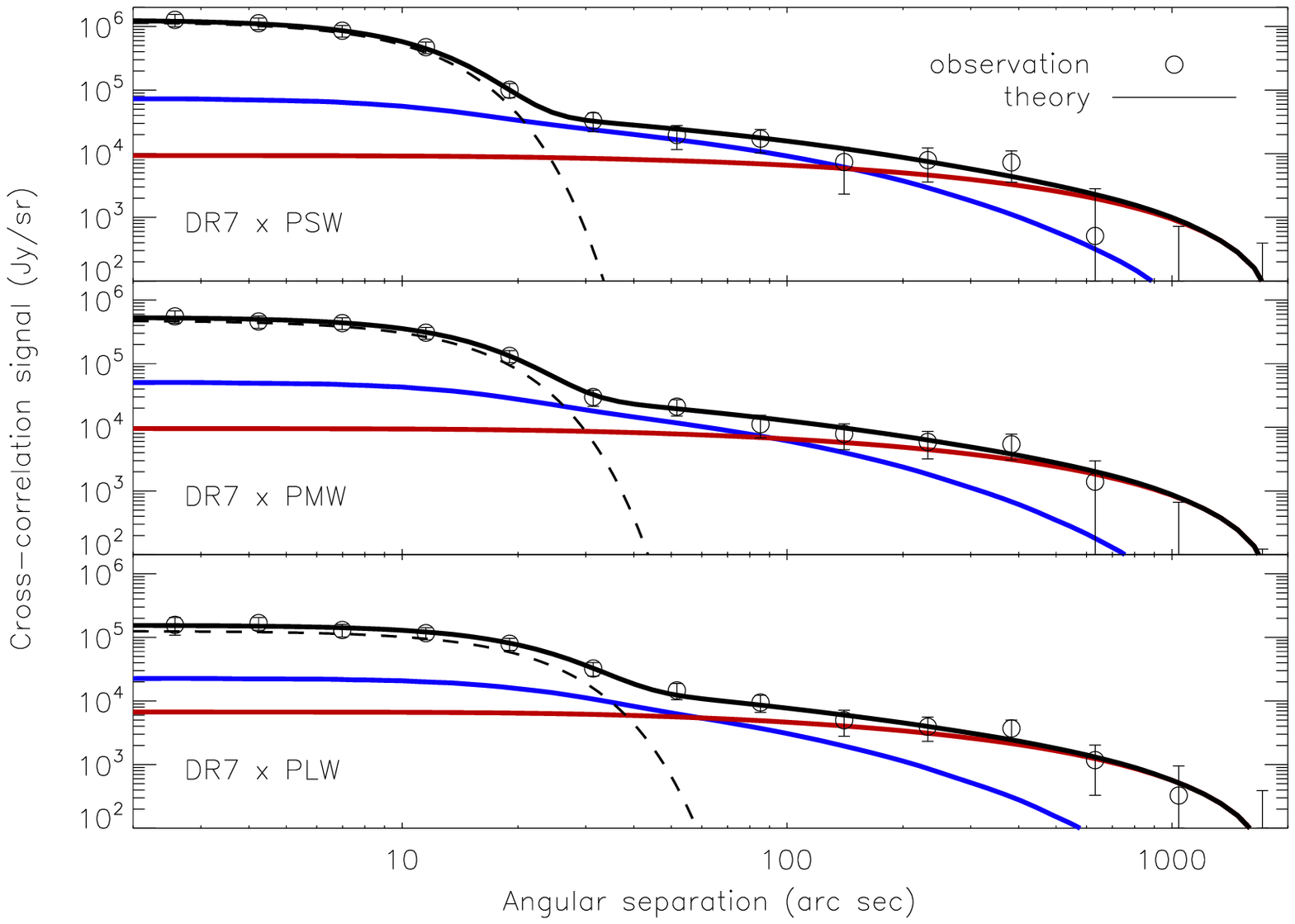}
\includegraphics[height=4.in,width=3.45in]{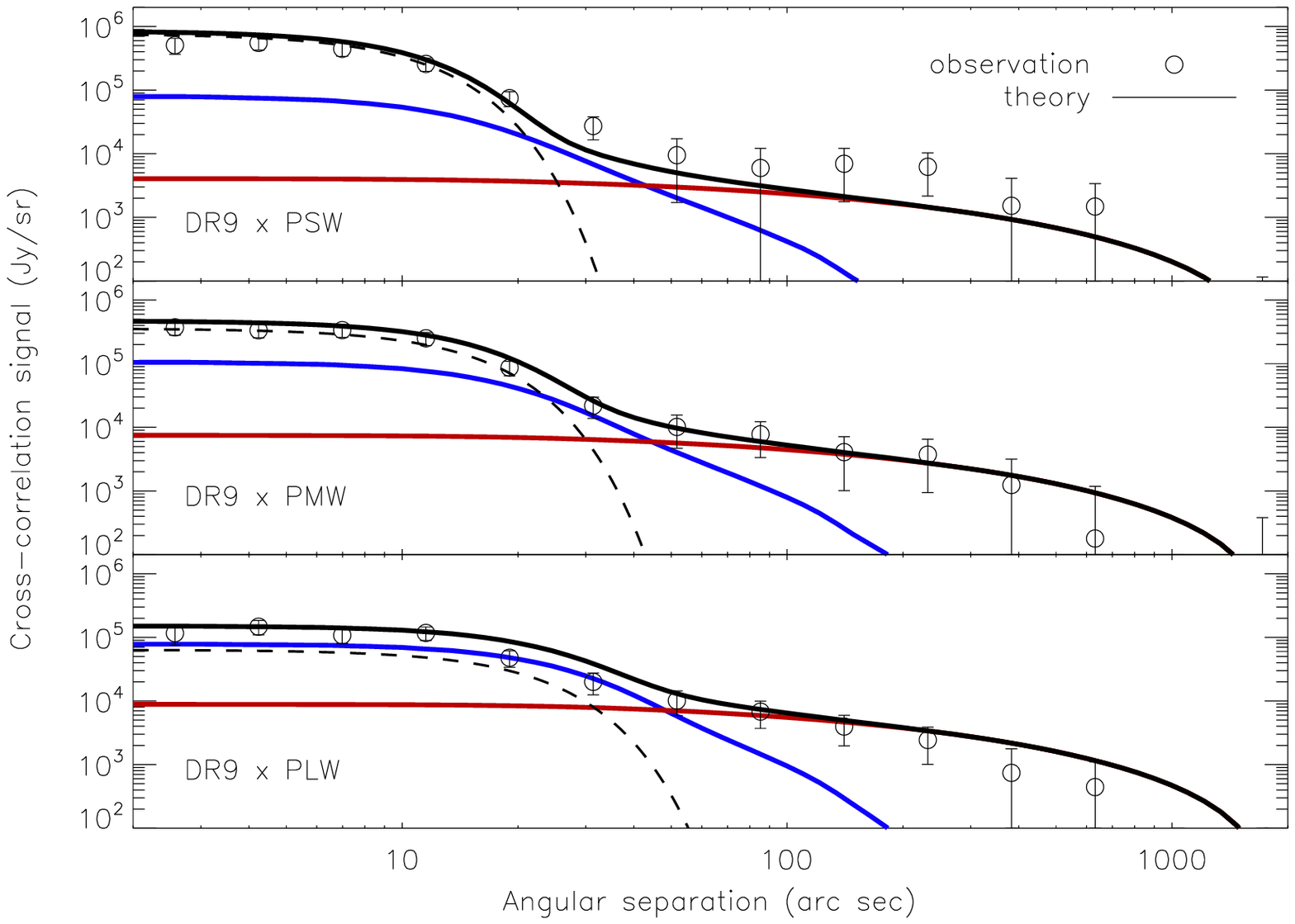}
\caption{The combined cross-correlation signal (over all fields) between the SDSS QSOs (left column: DR7; right column: DR9) and the CIB (top: 250\ $\micron$; middle: 350\ $\micron$; bottom: 500\ $\micron$) compared with the best-fit from the halo model. The dashed line in each panel corresponds to the best-fit mean sub-mm emission of the quasars themselves. The blue solid line  corresponds to the best-fit one-halo term of the cross-correlation signal between the QSOs and the CIB convolved with the 2D SPIRE beam, the red solid line corresponds to the best-fit two-halo term convolved with the 2D SPIRE beam. The black circles are the measured cross-correlation signal. The black solid line is the best-fit total signal from the halo model, i.e. the sum of the mean sub-mm emission of the quasars themselves and the correlated emission from DSFGs.}
\label{fig:stacked_signal_halomodel}
\end{figure*}

\subsection{Halo model of the cross-correlation between quasars and the CIB}

Combining the CIB halo model described in Section 4.1 and the quasar halo model described in Section 4.2, we can model the cross-correlation between the quasars and the CIB.
The 1-halo term of the cross-correlation power spectrum between quasars and the CIB can be modelled as 
\begin{eqnarray}
P_{sq}^{1h} (k, z) & = &\frac{1}{\bar{j}\bar{n}} \int dM \frac{dN}{dM} (z)  \\ \nonumber
 & & \times [f_s^{\rm{cen}}(M, z) N_q^{\rm{sat}}(M, z) u_{\rm{gal}}(k, z|M) + \\ \nonumber
  & & \times [f_s^{\rm{sat}}(M, z) N_q^{\rm{cen}}(M, z) u_{\rm{gal}}(k, z|M) + \\ \nonumber
 & & \times [f_s^{\rm{sat}}(M, z) N_q^{\rm{sat}}(M, z) u_{\rm{gal}}^2(k, z|M) ]
\end{eqnarray}
And the 2-halo term of the cross-correlation power spectrum between quasars and the CIB can be modelled as 
\begin{equation}
P_{sq}^{2h} (k, z) = \frac{1}{\bar{j}\bar{n}} P_{\rm{lin}}(k, z) D_s(k, z) D_q (k, z ),
\end{equation}
where
\begin{eqnarray}
D_s(k, z) & = &\int dM \frac{dN}{dM} (z) b(M, z) u_{\rm{gal}}(k, z|M) \times \\ \nonumber
 & & [f_s^{\rm{cen}}(M, z) + f_s^{\rm{sat}}(M, z)]
\end{eqnarray}
and
\begin{eqnarray}
D_q (k, z ) &= &\int dM \frac{dN}{dM} (z)b(M, z) u_{\rm{gal}}(k, z|M) \times \\ \nonumber
& &[N_q^{\rm{cen}}(M, z)  + N_q^{\rm{sat}}(M, z) ].
\end{eqnarray}

Under Limber's approximation (Limber 1953), the projected angular cross-correlation power spectrum between QSOs and CIB is related to the three-dimensional spatial power spectrum,
\begin{equation}
P^{\rm QS}(k_\theta) = \int \frac{dz}{dV_c/dz} P\left(k=\frac{k_{\theta}}{2\pi}, z\right) \frac{dS}{dz} \frac{dN^{\rm Q}}{dz},
\end{equation}
where $dV_c/dz$ is the comoving volume element per unit along the redshift axis, $dS/dz$ is the redshift distribution of the cumulative CIB flux, and $dN^{\rm Q}/dz$ is the normalised redshift distribution of the QSO sample.

The angular correlation function can be expressed in terms of the angular power spectrum
\begin{equation}
w(\theta) = \int \frac{k_{\theta} d k_{\theta}}{2\pi} P(k_\theta) J_0(k_{\theta}\theta),
\end{equation}
where $J_0(x)$ is the zeroth order Bessel function. Furthermore, the angular correlation function needs to be convolved with the 2D SPIRE beam, which can be described as a 2D Gaussian function. Therefore, the final beam convolved angular cross-correlation between the QSOs and the SPIRE maps can be written as
\begin{eqnarray}
w_{\rm conv}(\theta) &=& \frac{1}{2\pi \sigma^2} \int w(|\vec{\theta} - \vec{\theta'}|)  \exp\left(-\frac{\theta'^2}{2\sigma^2}\right) d\vec{\theta'} \nonumber \\
& = & \frac{1}{2\pi \sigma^2} \int_0^{2\pi} w(\sqrt{\theta^2 + \theta'^2 - 2\theta \theta'\cos{\phi}}) d\phi \nonumber \\ 
&  & \int\exp\left(-\frac{\theta'^2}{2\sigma^2}\right) \theta' d\theta', 
\end{eqnarray}
where $\sigma$ is the standard deviation of the Gaussian beam.

\subsection{Halo model results}

As mentioned earlier, we have used the same parameters of the CIB halo model as in Viero et al. (2013) which fits the auto- and cross-correlation power spectra of the SPIRE bands. So, in this paper we are only varying the parameters related to the quasar population to fit the cross-correlation signal between the quasars and the CIB. There are in total eight free parameters in our halo model, which are the five parameters describing the halo occupation distributions of the quasars and the three parameters related to the amplitude of the mean sub-mm emissions of the quasars themselves at 250, 350 and 500\ $\micron$.

In Fig.~\ref{fig:stacked_signal_halomodel}, we plot the combined cross-correlation signal, over all fields,  between the SDSS QSOs (left column: DR7 QSOs; right column: DR9 QSOs) and the CIB (top: 250\ $\micron$; middle: 350\ $\micron$; bottom: 500\ $\micron$) and the best-fit from the halo model. The dashed line in each panel corresponds to the best-fit mean sub-mm emission of the quasars themselves. The blue solid line  corresponds to the best-fit one-halo term of the cross-correlation signal convolved with the 2D SPIRE beam, the red solid line corresponds to the best-fit two-halo term convolved with the SPIRE beam, and the black solid line corresponds to the total cross-correlation signal (i.e. the sum of the sub-mm emission of the quasars themselves and the emission from the correlated DFSGs) convolved with the SPIRE beam. The black circles are the measured cross-correlation signal. With a median redshift $\left< z \right>$ of 1.4, the mean sub-mm flux densities of the DR7 quasars is $11.12^{+1.56}_{-1.37}$, $7.08^{+1.63}_{-1.33}$ and $3.63^{+1.35}_{-0.98}$ mJy at 250, 350 and 500\ $\micron$ respectively, while the mean sub-mm flux densities of the DR9 quasars ($\left< z \right>=2.5$) is $5.69^{+0.72}_{-0.64}$, $4.98^{+0.75}_{-0.65}$ and $1.76^{+0.50}_{-0.39}$ mJy at 250, 350 and 500\ $\micron$ respectively. Using a modified blackbody with a dust emissivity of 1.45 combined with the mid-infrared power-law ($f_{\nu}\propto \nu^{-2}$) to describe the SED of the quasars (i.e. the same assumptions which we made regarding the SED of the DSFGs in Section 4.1), we can derive the best-fit dust temperature and total infrared luminosity for the quasars. For the DR7 quasars, the best-fit dust temperature and infrared luminosity is $39.01^{+11.00}_{-6.25}$ K  and $\log_{10} L_{\rm IR} (L_{\odot}) = 12.38^{+0.25}_{-0.15} $ respectively. For the DR9 quasars, the best-fit dust temperature and infrared luminosity is $52.30^{+7.78}_{-5.56}$ K and $\log_{10} L_{\rm IR} (L_{\odot}) = 12.82^{+0.12}_{-0.09}$ respectively. Due to the degeneracy between dust temperature and dust emissivity, the best-fit temperature decreases with increasing values of dust emissivity but the best-fit infrared luminosity is almost unaffected. Therefore, the higher redshift SDSS-III DR9 quasars are on average brighter than the SDSS DR7 quasars in the infrared, possibly due to elevated star-formation activity and/or black hole accretion.  Another possibility is that for the higher redshift DR9 quasars, we are starting to probe shorter wavelength range where the contribution of the accreting BH to the total infrared luminosity is larger. The dust temperature of the satellite galaxies around DR7 and DR9 quasars is fixed at 24.2 K in the halo model (see Section 4.1) as we use the same CIB halo model parameters as  in Viero et al. (2013) which fits the auto- and cross-frequency channel power spectra at the SPIRE wavelengths. As we have used the same SED assumptions to describe the QSOs and the DSFGs, we can directly compare the estimated temperatures for both populations. Both SDSS DR7 and SDSS-III DR9 QSOs are found to be hotter than the correlated DSFGs, possibly due to hotter dust heated by the accreting BH.

\begin{figure}
\includegraphics[height=2.7in,width=3.45in]{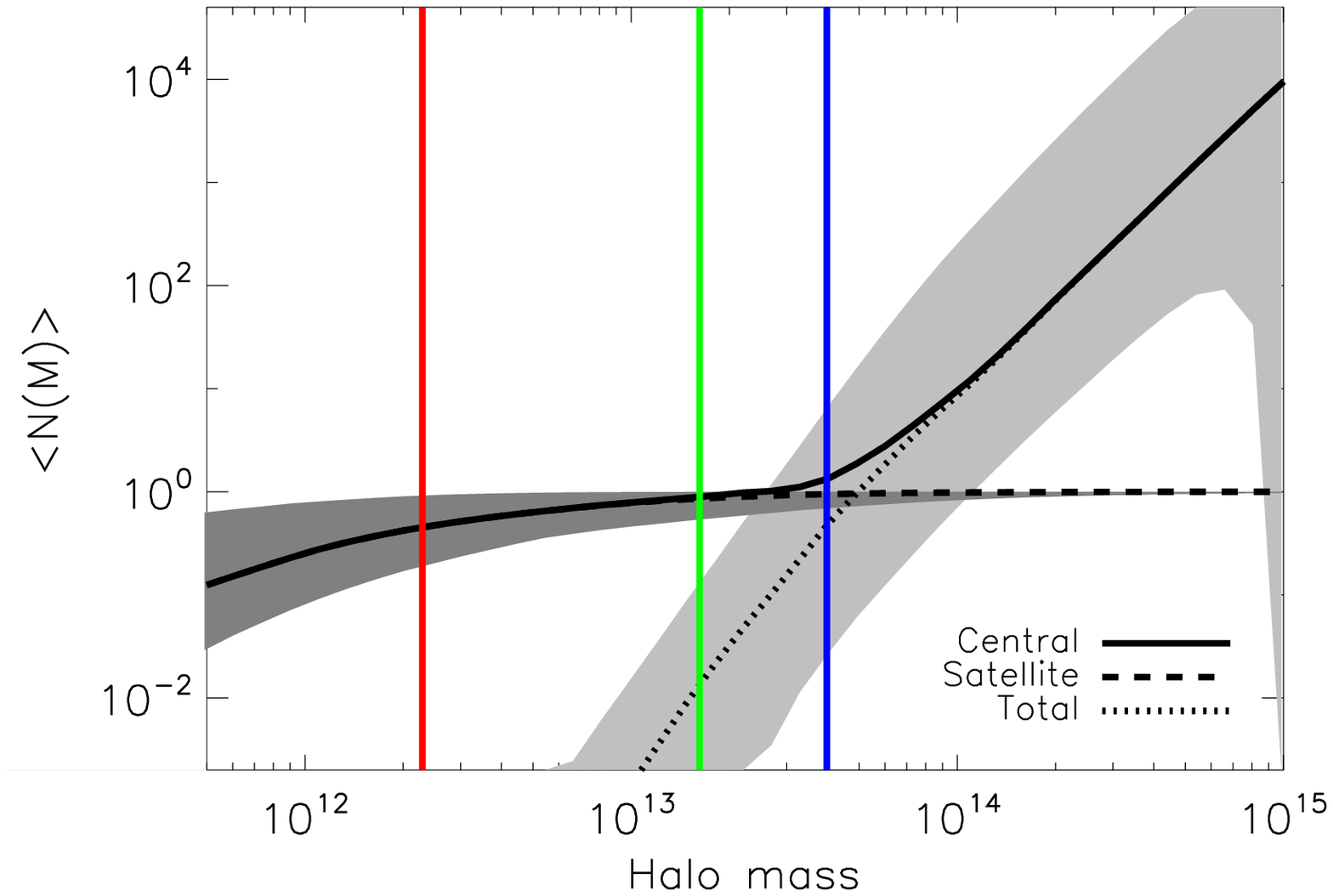}
\includegraphics[height=2.7in,width=3.45in]{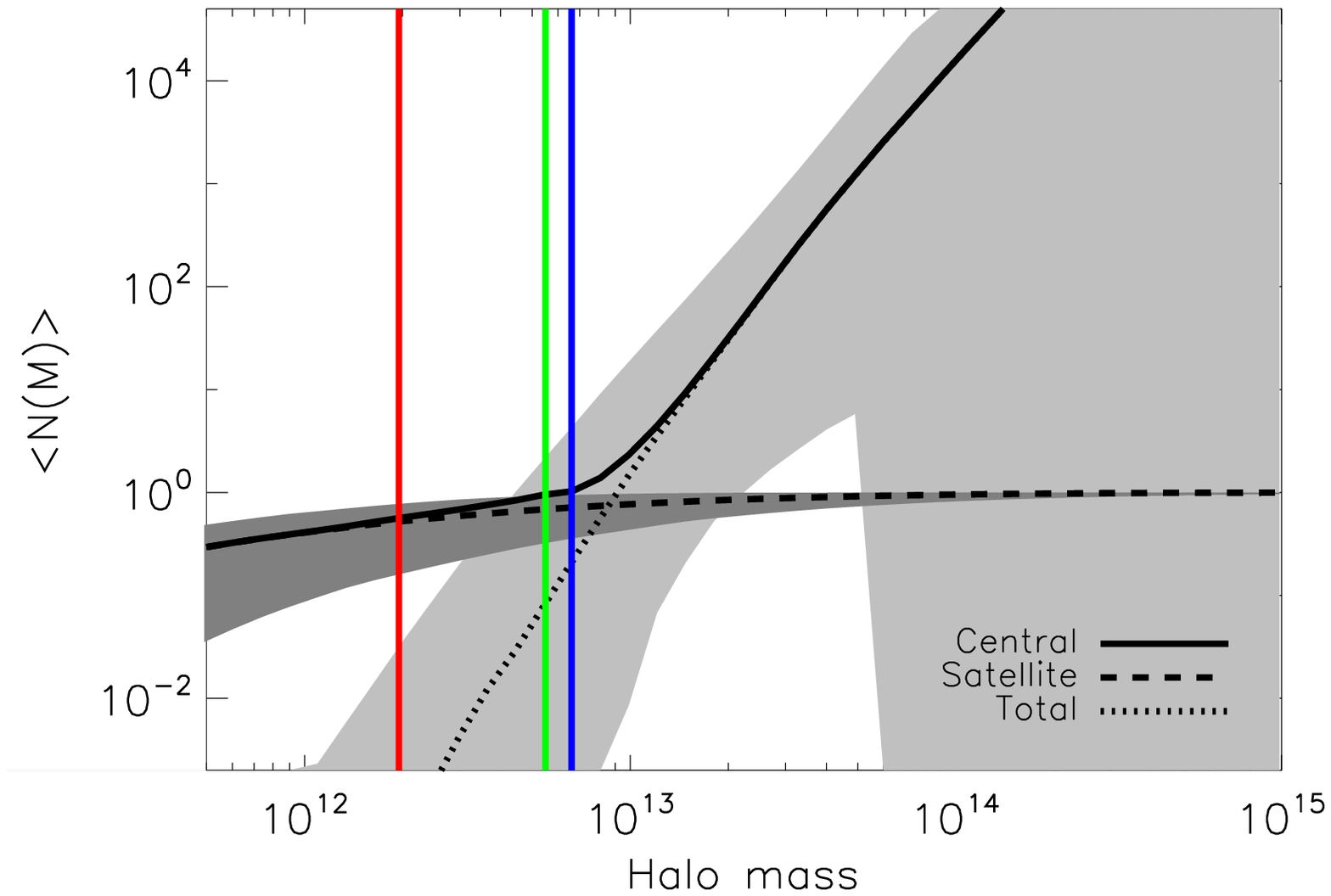}
\caption{The mean halo occupation distribution of SDSS QSOs (top: DR7; bottom: DR9), decomposed into its central and satellite components. The shaded region indicate the $68\%$ confidence intervals. The red, green and blue solid lines represent the best-fit value for $M_{\rm min}$ (the characteristic halo mass scale at which on average half of the halos host one central quasar), $M_{\rm cut}$ (the halo mass scale below which the HOD for the satellite quasars decays exponentially) and $M_{1}$ (the halo mass mass for hosting on average one satellite quasar) respectively. Note that the quasar HODs should be interpreted as the number of quasars in halos of mass $M$ which host quasars, not all halos of mass $M$ (see discussions on the quasar duty cycle in Section 4.2).}
\label{fig:hod_halomodel}
\end{figure}

In Fig.~\ref{fig:hod_halomodel}, we plot the mean halo occupation distribution of SDSS DR7 and DR9 QSOs, decomposed into its central and satellite components. The shaded region indicate the $68\%$ confidence intervals. The red, blue and green solid lines represent the best-fit value for $M_{\rm min}$ (the characteristic halo mass scale at which on average half of the halos host one central quasar), $M_{\rm cut}$ (the halo mass scale below which the HOD for the satellite quasars decays exponentially) and $M_{1}$ (the halo mass mass for hosting on average one satellite quasar) respectively. For the DR7 QSOs,  the host halo mass scale for the central and satellite quasars is $M_{\rm min}=10^{12.36\pm0.87}$ M$_{\odot}$ and $M_{1}=10^{13.60\pm0.38}$ M$_{\odot}$, respectively.  For the DR9 QSOs,  the host halo mass scale for the central and satellite quasars is $M_{\rm min}=10^{12.29\pm0.62}$ M$_{\odot}$ and $M_{1}=10^{12.82\pm0.39}$ M$_{\odot}$, respectively. The typical halo environment of the DR7 and DR9 QSOs we find in this study is similar to previous studies using QSO auto-correlation statistics (e.g., Shen et al. 2009; White et al. 2012; Shen et al. 2012; Richarson et al. 2012).Table 2 and Table 3 list the best-fit values, uncertainties and correlations of the parameters in the halo model for the DR7 and DR9 quasars.  From the HODs, we find that the satellite fraction (the ratio of satellite quasars to the total number of quasars) is $0.008^{+0.008}_{-0.005}$ and $0.065^{+0.021}_{-0.031}$ for the DR7 and DR9 quasars, respectively. Our estimates are broadly consistent with other studies in the literature, i.e. the QSO satellite fraction is found to be around a few percent at most. Thus the majority of the DR7 and DR9 QSOs are expected to be central galaxies in dark matter halos and the correlated DSFGs in the one-halo term are mostly satellite galaxies in the same halo.

\begin{table*}
\caption{Best-fit values, uncertainties and correlation of the model parameters for the SDSS DR7 QSOs. The first five parameters ($M_{\rm min}$, $\sigma_M$, $\alpha$, $M_1$ and $M_{\rm cut}$) describe the HODs of the DR7 QSOs while the last three parameters ($A_1$, $A_2$ and $A_3$) correspond to the mean sub-mm flux (in unit of Jy sr$^{-1}$) of the DR7 QSOs at 250, 350 and 500\ $\micron$, respectively. Using the SPIRE beam area, we can convert the mean sub-mm flux of the DR7 QSOs into $11.12^{+1.56}_{-1.37}$, $7.08^{+1.63}_{-1.33}$ and $3.63^{+1.35}_{-0.98}$ mJy at 250, 350 and 500\ $\micron$, respectively.}
    \begin{tabular}{ccccccccc}
    \hline
    Parameter & $\log_{10}M_{\rm min}$ & $\sigma_M$ & $\alpha$ & $\log_{10}M_1$ & $\log_{10}M_{\rm cut}$ & $\log_{10} A_1$ & $\log_{10} A_2$ & $\log_{10} A_3$\\
    \hline
      $M_{\rm min}$ & $12.36\pm0.87 $ & 0.13  & -0.08     &0.72 &  -0.10 &   -0.41 &   -0.44   & -0.45 \\
     $\sigma_M$      & -- & $1.12\pm0.52$ & 0.32  & -0.27     &0.20    & 0.36     &0.43     &0.47\\
     $\alpha$            & -- &--& $2.97\pm0.51$ & 0.19  &-0.10    & 0.21     &0.29     &0.30\\
      $M_1$              &-- &--&--& $13.60\pm0.38$ & -0.44 &-0.61   & -0.65    &-0.69\\
    $M_{\rm cut}$    & --&--&--&--& $13.21\pm0.49$ & 0.29   &0.36   &  0.39 \\
     $A_1$                &-- &--&--&--&--& $6.05\pm0.06$ &0.62   &0.69\\
      $A_2$               &-- &--&--&--&--&--&$5.60\pm0.09$&0.79\\
      $A_3$               & --&--&--&--&--&--&--&$4.99\pm0.14$\\
    \hline
    \end{tabular}
\end{table*}

\begin{table*}
\caption{Best-fit values, uncertainties and correlation of the model parameters for the SDSS-III DR9 QSOs. The first five parameters ($M_{\rm min}$, $\sigma_M$, $\alpha$, $M_1$ and $M_{\rm cut}$) describe the HODs of the DR9 QSOs while the last three parameters ($A_1$, $A_2$ and $A_3$) correspond to the mean sub-mm flux  (in unit of Jy sr$^{-1}$) of the DR9 QSOs at 250, 350 and 500\ $\micron$, respectively. Using the SPIRE beam area, we can convert the mean sub-mm flux of the DR9 QSOs into $5.69^{+0.72}_{-0.64}$, $4.98^{+0.75}_{-0.65}$ and $1.76^{+0.50}_{-0.39}$ mJy at 250, 350 and 500\ $\micron$, respectively.}
    \begin{tabular}{ccccccccc}
    \hline
    Parameter & $\log_{10}M_{\rm min}$ & $\sigma_M$ & $\alpha$ & $\log_{10}M_1$ & $\log_{10}M_{\rm cut}$ & $\log_{10} A_1$ & $ \log_{10} A_2$ & $\log_{10} A_3$\\
    \hline
      $M_{\rm min}$ & $12.29\pm0.62 $ &  0.24 & -0.38 &  0.52 &   -0.30  & -0.02 &  -0.08 &  -0.10\\
     $\sigma_M$      & -- & $1.33\pm0.54$ &  0.20 & -0.49 & -0.21 &  -0.05 &   -0.18 &   -0.26\\
     $\alpha$            & -- &--& $3.21\pm0.61$ & -0.22  &   0.19&    0.06 &   0.04 &    0.05\\
      $M_1$              &-- &--&--& $12.82\pm0.39$ & -0.25 &   0.10 &    0.10  &   0.19\\
    $M_{\rm cut}$    & --&--&--&--& $12.74\pm0.61$ &  0.13  & 0.29&   0.35 \\
     $A_1$                &-- &--&--&--&--& $5.76\pm0.05$ &0.21  &   0.20\\
      $A_2$               &-- &--&--&--&--&--&$5.45\pm0.06$& 0.49\\
      $A_3$               & --&--&--&--&--&--&--&$4.67\pm0.11$\\
    \hline
    \end{tabular}
\end{table*}

\section{Discussions and conclusions}

We present the first cross-correlation measurement between optically selected Type 1 SDSS DR7 and DR9 quasars and the cosmic infrared background (CIB) measured by the {\em Herschel}-SPIRE instrument at 250, 350 and 500\ $\micron$. The distribution of the quasars  at $0.15<z<3.5$ (SDSS DR7 quasars: $0.15<z<2.5$; SDSS-III DR9 quasars: $2.2<z<3.5$) covers the redshift range where we expect most of the CIB to originate. We detect the sub-millimetre (sub-mm) emission of the optical quasars which dominates on small angular scales as well as the correlated sub-mm emission from dusty star-formation galaxies (DSFGs) dominant on larger angular scales.

A simple halo model is used to interpret the measured cross-correlation signal between the quasars and the CIB. At a median redshift of 1.4, the mean sub-mm flux densities  of the DR7 quasars is $11.12^{+1.56}_{-1.37}$, $7.08^{+1.63}_{-1.33}$ and $3.63^{+1.35}_{-0.98}$ mJy at 250, 350 and 500\ $\micron$ respectively. At a median redshift of 2.5, the mean sub-mm flux densities of the DR9 quasars is $5.69^{+0.72}_{-0.64}$, $4.98^{+0.75}_{-0.65}$ and $1.76^{+0.50}_{-0.39}$  mJy at 250, 350 and 500\ $\micron$ respectively.  By fitting a modified blackbody SED combined with a mid-infrared power-law template to the mean sub-mm fluxes, we find that the higher-redshift DR9 quasars have a higher mean infrared luminosity than the DR7 quasars, possibly due to elevated star-formation activity and/or black hole accretion or a larger contribution to the infrared luminosity from the accreting BH. Further investigations into the power source of the infrared luminosity of the quasars using data at other wavelengths is beyond the scope of this paper. We find that the correlated sub-mm emission  includes both the emission from satellite DSFGs in the same dark matter halo as the central quasar (the one-halo term) and the emission from DSFGs in separate halos which are correlated with the quasar-hosting halo (the two-halo term). The amplitude of the one-halo term of the correlated emission is about 10 times smaller than the sub-mm emission of the quasars themselves, implying that the satellite dusty star-forming galaxies are much fainter and have a lower SFR than the central quasars. 

We infer from the halo model that the satellite fraction of the SDSS DR7 quasars ($z=[0.15, 2.5]$) at a median redshift $\left< z \right>$ of 1.4 is $0.008^{+0.008}_{-0.005}$, and the host halo mass scale for the central and satellite quasars is $M_{\rm min}=10^{12.36\pm0.87}$ M$_{\odot}$  and $M_{1}=10^{13.60\pm0.38}$ M$_{\odot}$ respectively. The satellite fraction of the SDSS-III DR9 quasars ($z=[2.2, 3.5]$)  at a median redshift $\left< z \right>$  of 2.5 is $0.065^{+0.021}_{-0.031}$ , and the host halo mass scale for the central and satellite quasars is $M_{\rm min}=10^{12.29\pm0.62}$ M$_{\odot}$ and  $M_{1}=10^{12.82\pm0.39}$ M$_{\odot}$ respectively. Therefore, the typical halo environment of the quasars is similar to that of the DSFGs over the redshift range probed. This is expected if dusty starburst and quasar activity are evolutionarily linked phenomena.

Several aspects of our study could be improved and/or extended. In this paper, we have used the cross-correlation signal between the quasar samples and the CIB maps to constrain the halo occupation statistics of the quasars. A more complete approach would be to combine the auto-correlation statistics (of the quasars and the CIB maps) and the cross-correlation statistics (between the two tracers) to constrain the connection between different types of galaxies and the underlying dark matter halos simultaneously. We leave this project to a future paper when we have a better understanding of the QSO selection effects in the Stripe 82 region. Donoso et al. (2013) recently reported a difference in the halo environment between obscured and unobscured AGN. We can provide independent constraints by looking at the cross-correlation between obscured and unobscured AGN and the CIB (Wang et al. 2014, in prep.).

\section*{ACKNOWLEDGEMENTS}
LW and PN acknowledge support from an ERC StG grant (DEGAS-259586). 

SPIRE has been developed by a consortium of institutes led by Cardiff Univ. (UK) and including Univ. Lethbridge (Canada); NAOC (China); CEA, LAM (France); IFSI, Univ. Padua (Italy); IAC (Spain); Stockholm Observatory (Sweden); Imperial College London, RAL, UCL-MSSL, UKATC, Univ. Sussex (UK); Caltech, JPL, NHSC, Univ. Colorado (USA). This development has been supported by national funding agencies: CSA (Canada); NAOC (China); CEA, CNES, CNRS (France); ASI (Italy); MCINN (Spain); SNSB (Sweden); STFC (UK); and NASA (USA).

Funding for the SDSS and SDSS-II has been provided by the Alfred P. Sloan Foundation, the Participating Institutions, the National Science Foundation, the U.S. Department of Energy, the National Aeronautics and Space Administration, the Japanese Monbukagakusho, the Max Planck Society, and the Higher Education Funding Council for England. The SDSS Web Site is http://www.sdss.org/.

The SDSS is managed by the Astrophysical Research Consortium for the Participating Institutions. The Participating Institutions are the American Museum of Natural History, Astrophysical Institute Potsdam, University of Basel, University of Cambridge, Case Western Reserve University, University of Chicago, Drexel University, Fermilab, the Institute for Advanced Study, the Japan Participation Group, Johns Hopkins University, the Joint Institute for Nuclear Astrophysics, the Kavli Institute for Particle Astrophysics and Cosmology, the Korean Scientist Group, the Chinese Academy of Sciences (LAMOST), Los Alamos National Laboratory, the Max-Planck-Institute for Astronomy (MPIA), the Max-Planck-Institute for Astrophysics (MPA), New Mexico State University, Ohio State University, University of Pittsburgh, University of Portsmouth, Princeton University, the United States Naval Observatory, and the University of Washington.

Funding for SDSS-III has been provided by the Alfred P. Sloan Foundation, the Participating Institutions, the National Science Foundation, and the U.S. Department of Energy Office of Science. The SDSS-III web site is http://www.sdss3.org/.

SDSS-III is managed by the Astrophysical Research Consortium for the Participating Institutions of the SDSS-III Collaboration including the University of Arizona, the Brazilian Participation Group, Brookhaven National Laboratory, Carnegie Mellon University, University of Florida, the French Participation Group, the German Participation Group, Harvard University, the Instituto de Astrofisica de Canarias, the Michigan State/Notre Dame/JINA Participation Group, Johns Hopkins University, Lawrence Berkeley National Laboratory, Max Planck Institute for Astrophysics, Max Planck Institute for Extraterrestrial Physics, New Mexico State University, New York University, Ohio State University, Pennsylvania State University, University of Portsmouth, Princeton University, the Spanish Participation Group, University of Tokyo, University of Utah, Vanderbilt University, University of Virginia, University of Washington, and Yale University.

\appendix

\section{IRIS 100\ $\micron$ maps}
\label{appendix1}

\begin{figure*}
\includegraphics[height=2.in,width=7in]{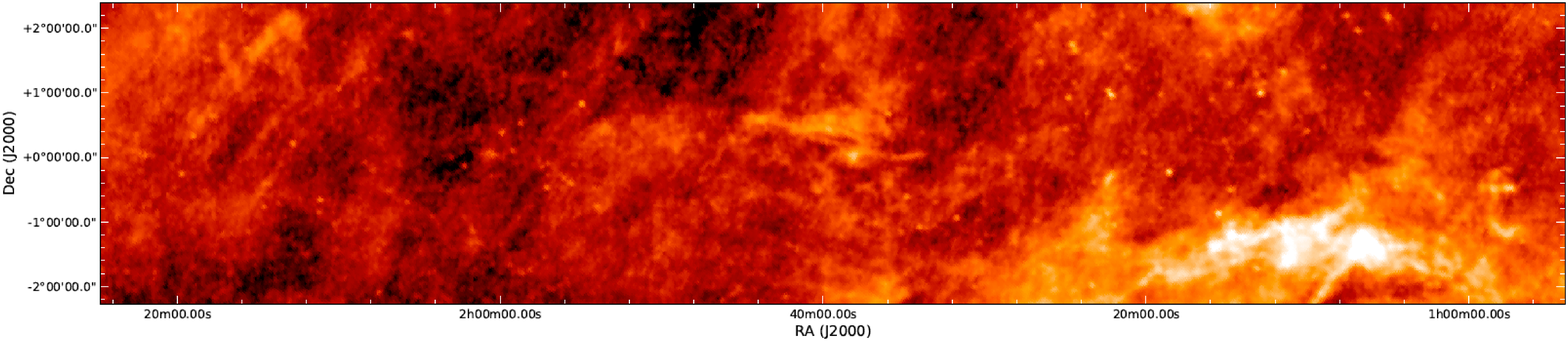}
\includegraphics[height=2.in,width=7in]{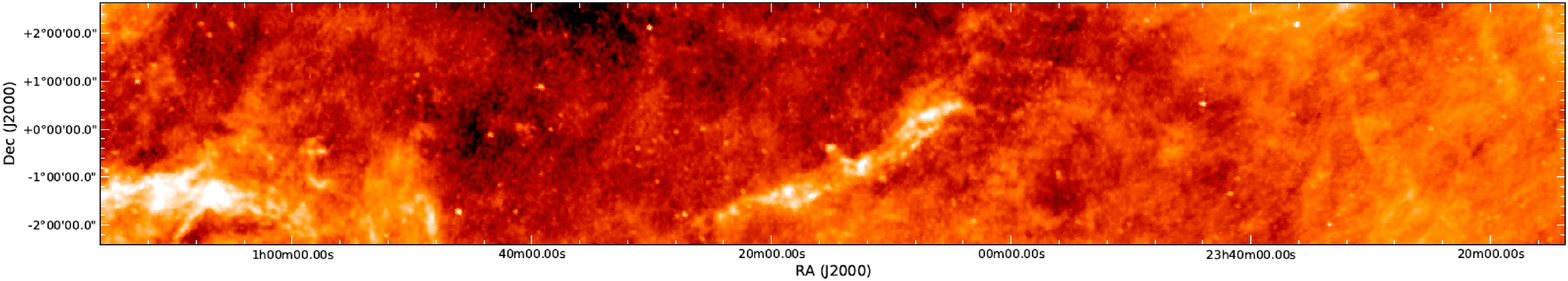}
\caption{The IRIS 100\ $\micron$ maps in the HeRS (top) and HeLMS (bottom) field. The dominant signal in these maps comes from extended diffuse Galactic dust emission.}
\label{fig:iris100_hershelms}
\end{figure*}

\begin{figure}
\includegraphics[height=3.05in,width=3.55in]{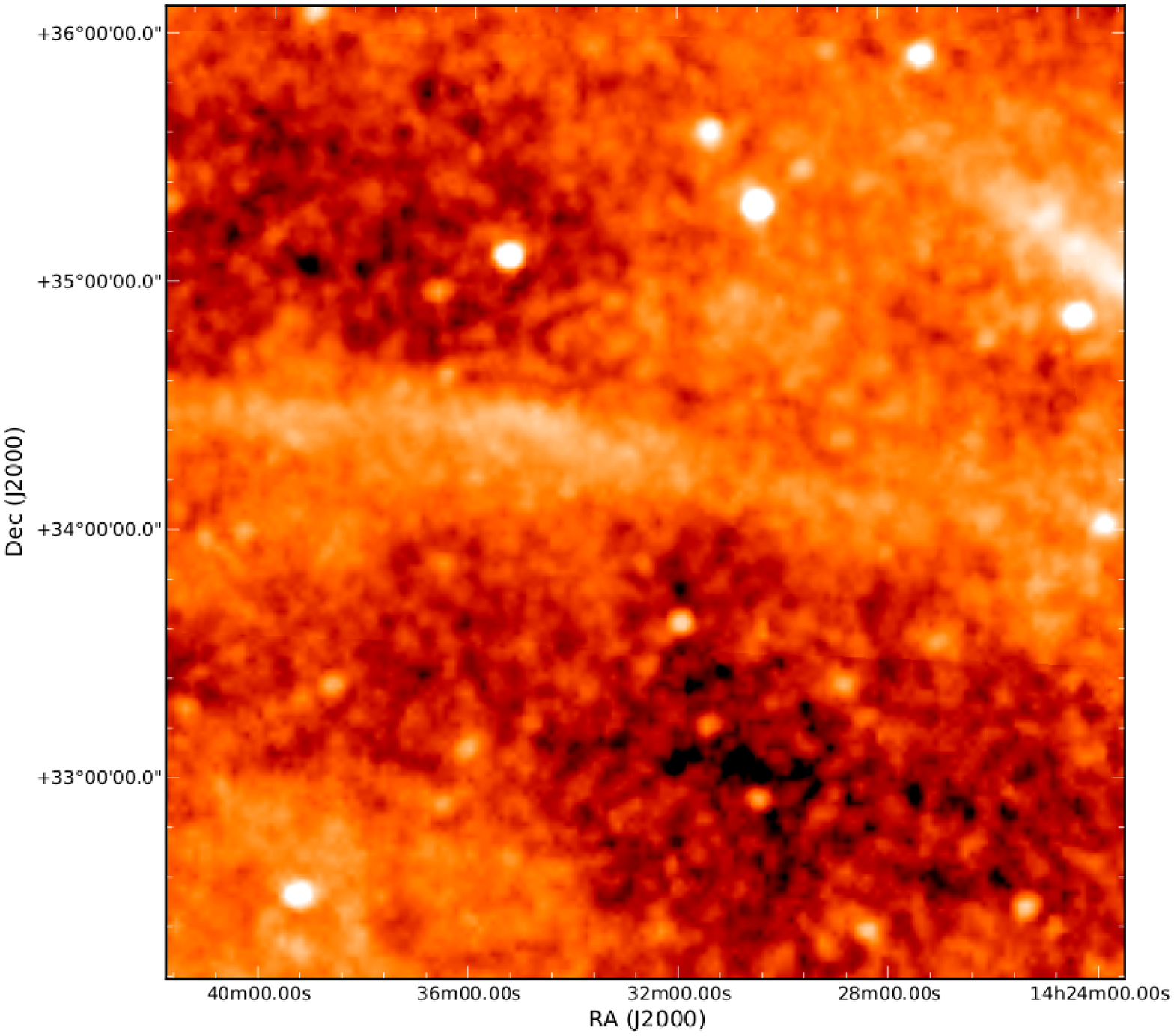}
\includegraphics[height=3.05in,width=3.55in]{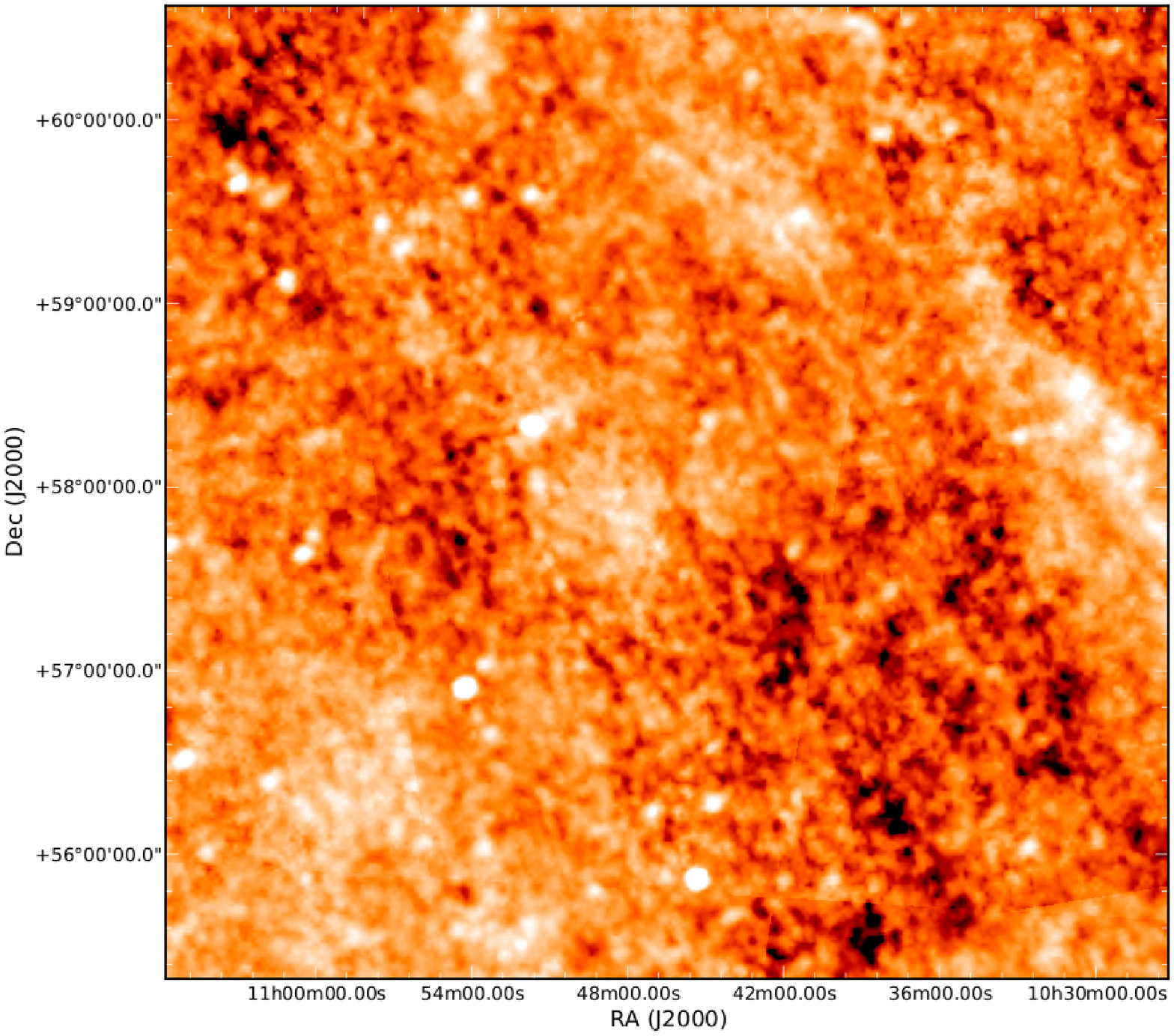}
\includegraphics[height=3.05in,width=3.55in]{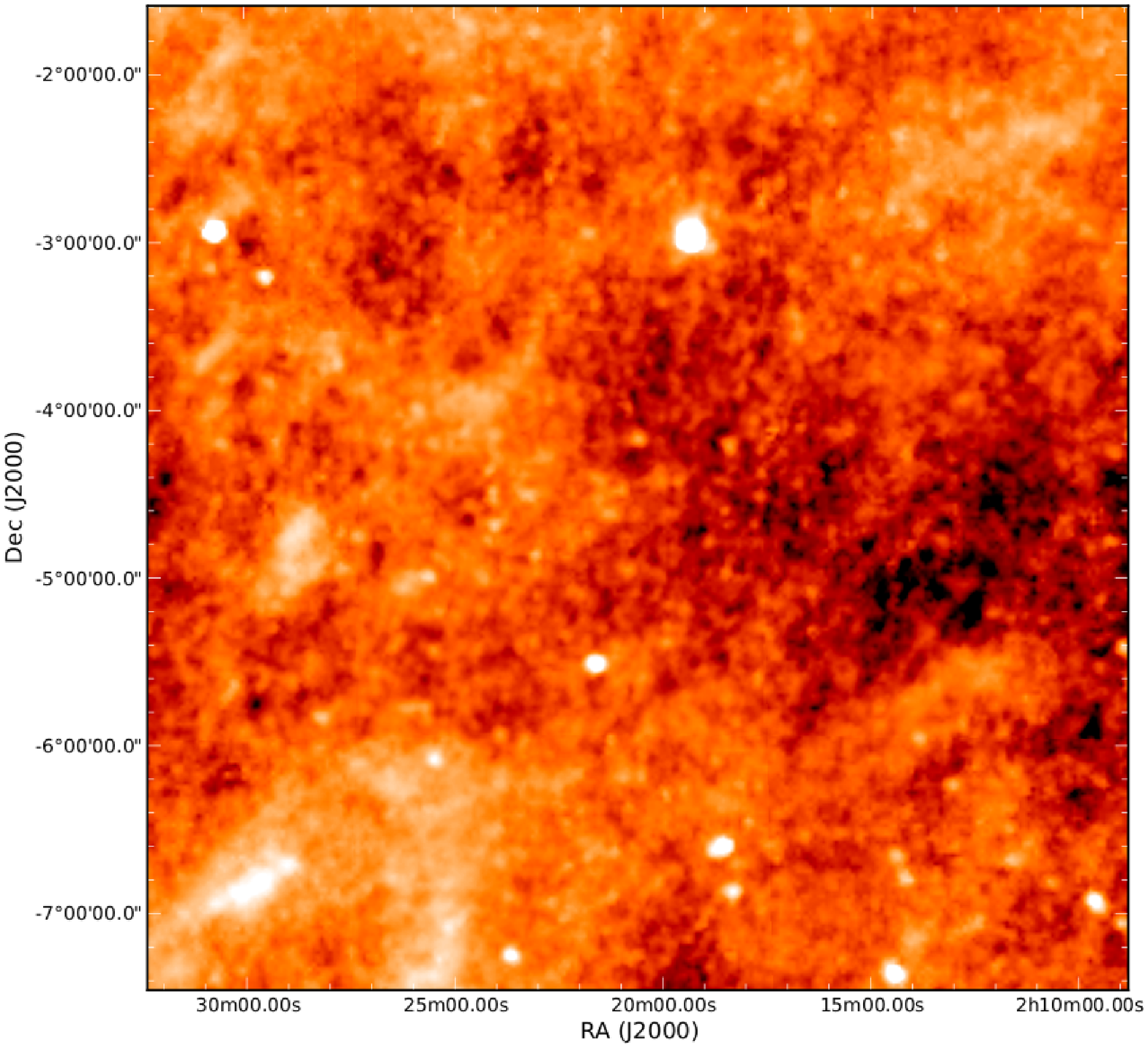}
\caption{The IRIS 100\ $\micron$ maps in the Bo\"{o}tes (top), Lockman-SWIRE (middle) and {\it XMM}-LSS (bottom) field.}
\label{fig:iris100_lss14}
\end{figure}

The IRIS 100\ $\micron$ maps in the HeRS and HeLMS field are plotted in Fig.~\ref{fig:iris100_hershelms}. The IRIS 100\ $\micron$ maps in the Bo\"{o}tes, Lockman-SWIRE and {\it XMM}-LSS field are plotted in Fig.~\ref{fig:iris100_lss14}.  Visual inspection confirms that extended diffuse Galactic cirrus emission are the dominant structures (long filamentary chains) in these 100\ $\micron$ maps, especially in the HeRS and HeLMS regions. The mean value of the 100\ $\micron$ maps in the HeLMS, HeRS, {\it XMM}-LSS, Bo\"{o}tes and Lockman-SWIRE field is  3.03, 2.84, 2.30, 1.42 and 1.09 MJy sr$^{-1}$, respectively. Adopting the DIRBE measurement  of the cosmic IR background at 100\ $\micron$ which is $14.4\pm6.3$ nW m$^{-2}$ sr$^{-1}$ or equivalently $0.48\pm0.21$ MJy  sr$^{-1}$ (Dole et al. 2006), we can estimate that the fractional contribution of the CIB to the 100\ $\micron$ maps in the HeLMS, HeRS, {\it XMM}-LSS, Bo\"{o}tes and Lockman-SWIRE field  is  16\%, 17\%, 21\%, 34\% and 44\%, respectively.

\section{Stacking}
\label{appendix2}

We have a quasar catalogue from which build the quasar density map $Q$, which gives the number of quasars in each pixel of the map. The mean value of this map is $\bar{Q} = N_\mathrm{obj} / N_\mathrm{pix}$, where $N_\mathrm{obj}$ is the number of objects in the catalogue and $N_\mathrm{pix}$ is the number of pixels in the map. The SPIRE map $M$ has a mean value $\bar{M}$ which is set to zero, as SPIRE has no sensitivity to the absolute zero-level on the sky. We retain the mean value in what follows to keep the equations general.

The covariance of $Q$ and $M$ is defined as:
\begin{eqnarray}
\label{eqn:cov}
\mathrm{Cov}(Q,M) &= & \langle (Q - \bar{Q}) (M - \bar{M}) \rangle  \nonumber \\ 
&=&\frac{1}{N_\mathrm{pix}} \sum_{i^\prime=1}^{N_\mathrm{x}} \sum_{j^\prime=1}^{N_\mathrm{y}} (Q_{i^\prime j^\prime} - \bar{Q}) (M_{i^\prime j^\prime} - \bar{M}),
\end{eqnarray}
where the sums are over the two map dimensions and $N_\mathrm{pix} =N_\mathrm{x} N_\mathrm{y}$. As shown in Marsden et al.\ (2009), this is equivalent to what is often called a ``stack'', when normalised by $\bar{Q}$, the average number of quasars per pixel:
\begin{equation}
\label{eqn:stack}
S = \frac{1}{\bar{Q}} \, \mathrm{Cov}(Q,M).
\end{equation}
We can extend Eq.~\ref{eqn:cov} to describe a stacked thumbnail, made by making a pixel-by-pixel average of map cutouts centred on each source. This is equivalent to calculating the covariance of two maps with a relative shift by $i,j$ pixels:
\begin{eqnarray}
\label{eqn:covij}
\mathrm{Cov}_{ij}(Q,M) &=& \frac{1}{(N_\mathrm{x}-i) (N_\mathrm{y}-j)} \sum_{i^\prime} \sum_{j^\prime} \nonumber \\
&&(Q_{i^\prime j^\prime} - \bar{Q}) (M_{(i^\prime+i)(j^\prime+j)} - \bar{M}).
\end{eqnarray}
In order to account for the edges of the maps, the sum over $i^\prime$($j^\prime$) runs from the larger of 1 and $i$ ($j$) to the smaller of
$N_\mathrm{x}$ ($N_\mathrm{y}$) and $N_\mathrm{x} - i$ ($N_\mathrm{y}- j$). $\mathrm{Cov}_{ij}(Q,M)$ is thus a map with indices running
from $-N_\mathrm{thumb}/2$ to $N_\mathrm{thumb}/2$ for an $N_\mathrm{thumb} \times N_\mathrm{thumb}$ thumbnail map. We can then
take an azimuthal average of this map to create the radial profile of the stack:
\begin{equation}
\label{eqn:rk}
R_k(Q,M) = \frac{1}{N_\mathrm{k}} \sum_{i,j \in \xi_k} \mathrm{Cov}_{ij}(Q,M),
\end{equation}
where $\xi_k$ is the set of pixels for which:
\begin{equation}
k \le \sqrt{i^2 + j^2} < k+1,
\end{equation}
and $N_\mathrm{k}$ is the number of pixels in $\xi_k$. Using Eq.~\ref{eqn:stack}, the radial stack $S_\theta$ is then:
\begin{equation}
\label{eqn:radstack}
S_\theta = \frac{1}{\bar{Q}} \, R_k(Q,M).
\end{equation}

In Section 3.1, we use a modification of the Landy \& Szalay (1993) two-point correlation function estimator,
\begin{equation}
\label{eqn:wtheta}
w(\theta) = \frac{D_1 D_2(\theta) - D_1 R_1(\theta) - R_1 D_1(\theta)
  + R_1 R_2(\theta)}{R_1 R_2(\theta)}.
\end{equation}
Here, $D_1$ and $D_2$ represent the quasar density map and SPIRE map, called $Q$ and $M$ above. $R_1$ and $R_2$ represent random realisations of these two datasets. The quantities $X_1 X_2(\theta)$ are the same as $R_k(X_1,X_2)$ in Eq.~\ref{eqn:rk} above, although without the subtractions of the map means in Eq.~\ref{eqn:covij}; for convenience, we refer to the non--mean-subtracted version of Eq.~\ref{eqn:rk} as $\tilde{R}_k(X_1,X_2)$. The first term in Eq.~\ref{eqn:wtheta} can then be precisely defined as $\tilde{R}_k(Q,M)$. The remaining terms all involve one or more random distributions, and thus in each term, the two quantities are uncorrelated. For uncorrelated distributions, the expectation value of the product of the two variables is equal to the product of the expectation values of each variable, $\langle x y \rangle = \langle x \rangle \langle y \rangle$, and so the remaining terms are simply to the product of the mean values, $\bar{Q} \bar{M}$. We can thus simplify Eq.~\ref{eqn:wtheta}:
\begin{equation}
w(\theta) = \frac{\tilde{R}_k(Q,M) - \bar{Q}\bar{M}}{\bar{Q}\bar{M}}.
\end{equation}
By expanding Eqn~\ref{eqn:covij}, Eq.~\ref{eqn:radstack} can
similarly be rewritten in terms of $\tilde{R}_k$:
\begin{equation}
S_\theta = \frac{1}{\bar{Q}} \left[ \tilde{R}_k(Q,M) - \bar{Q}\bar{M} \right].
\end{equation}
The denominator of Eq.~\ref{eqn:wtheta} is zero when the SPIRE map mean is zero, and so we modify the equation by multiplying both sides by the mean $\bar{M}$. We see that the algorithm used in Section 3.1 is thus exactly equivalent to Eq.~\ref{eqn:radstack}, the azimuthally-averaged stacked thumbnail image.


\end{document}